\documentclass[pra, a4paper, twocolumn, 10pt, showpacs]{revtex4}
\usepackage{bbm, amsmath, amssymb, amsthm, bm,textcomp, nicefrac}
\usepackage[T1]{fontenc}
\usepackage[latin9]{inputenc}
\usepackage{graphicx}
\usepackage{geometry}

\newcommand{\ket}[1]{\left|{#1}\right\rangle}
\newcommand{\bra}[1]{\left\langle{#1}\right|}
\newcommand{\braket}[2]{\langle{#1}|{#2}\rangle}
\newcommand{\braketd}[1]{\langle{#1}|{#1}\rangle}
\newcommand{\ketbrad}[1]{\left|{#1}\rangle\!\langle{#1}\right|}
\newcommand{\ketbra}[2]{\left|{#1}\rangle\!\langle{#2}\right|}
\newcommand{\EV}[2]{\langle{#1}\rangle_{#2}}

\begin{document}

\title{Tensor operators: constructions and applications for long-range interaction systems}

\author{F.\ Fr\"owis$^{1}$, V.\ Nebendahl$^{1}$ and W.\ D\"ur$^{1}$ }

\affiliation{$^1$ Institut f\"ur Theoretische Physik, Universit\"at
  Innsbruck, Technikerstr. 25, A-6020 Innsbruck,
  Austria}
\date{\today}

\begin{abstract}
We consider the representation of operators in terms of tensor networks and their application to ground-state approximation and time evolution of systems with long-range interactions. We provide an explicit construction to represent an arbitrary many-body Hamilton operator in terms of a one-dimensional tensor network, i.e. as a matrix product operator. For pairwise interactions, we show that such a representation is always efficient and requires a tensor dimension growing only linearly with the number of particles. For systems obeying certain symmetries or restrictions we find optimal representations with minimal tensor dimension. We discuss the analytic and numerical approximation of operators in terms of low-dimensional tensor operators. We demonstrate applications for time evolution and ground-state approximation, in particular for long-range interaction with inhomogeneous couplings. The operator representations are also generalized to other geometries such as trees and 2D lattices, where we show how to obtain and use efficient tensor network representations respecting a given geometry. 
\end{abstract}


\maketitle


\section{Introduction}\label{sec:intro}

The description of quantum systems in terms of tensor networks has attracted increased attention in recent years. Based on such a description, numerical and analytical methods to treat strongly correlated quantum systems have been put forward, where matrix product states (MPS) \cite{BaxterMPS,Baxter,Schadschneider,MPS,PGVWC07} used within the density matrix renormalization group (DMRG) \cite{DMRG}, projected entanglement pair states (PEPS) \cite{VC04,MVC07,Nishino2D} and the multiscale entanglement renormalization ansatz (MERA) \cite{MERA} can be mentioned as prominent examples. The common idea of these approaches is to represent the state of a quantum system in terms of a tensor network of low-rank tensors with a small dimension. While a generic quantum state of $N$ particles is described by a rank $N$ tensor, i.e. by exponentially many parameters, the number of parameters required to describe a network of low-rank tensors with small dimension is low. One hence obtains a subset of quantum states that can be efficiently described in this way, where the choice of the geometry of the tensor networks determines the (entanglement) features of the corresponding states and their possible relevance to describe quantum states of interest, e.g. ground states of strongly correlated quantum systems with a given geometry. For example, MPS and PEPS correspond to the choice of a 1D or 2D tensor network respectively, and turned out to be capable of efficiently describing a wide range of ground states of 1D or 2D quantum systems \cite{GroundsState}. Notice that the tensor network has to be contracted in order to determine relevant quantities such as coefficients of the state, its norm or expectation values of observables, and the possibility to efficiently contract the network in an approximate way is required for practical applications and numerical simulations. For these contractions, the tensor dimension $D$ plays a crucial role and determines the efficiency of the algorithms. Only relatively small values of $D$ can be handled in practice.

It is natural to apply a similar approach to describe operators rather than state vectors in terms of tensor networks. This has been implicitly done in \cite{X96} in the context of momentum space DMRG and formally initiated in \cite{CB08, McC07, McC08, MCPV08, CDV08}, where matrix product operator descriptions corresponding to one-dimensional tensor networks have been introduced and studied. The advantage of such an approach lies in the possibility to describe operators in a compact and efficient way, and to evaluate quantities of interest such as the expectation value of an operator (e.g. the Hamiltonian of a system) more efficiently. Rather than considering each interaction term in the Hamiltonian individually, leading to multiple contractions, the usage of a tensor network description of the Hamiltonian allows for the evaluation of the expectation value of the whole Hamiltonian in a single run. Furthermore, the properties of the operators can be systematically studied and related to entanglement features. Again, the efficiency of the corresponding algorithms depend on the tensor dimension $D$, and hence an optimized representation of the tensor network with low tensor dimension $D$ is desirable.  

In this paper we study systematic ways to construct such tensor network descriptions of arbitrary operators using linear tensor networks, so-called matrix product operators (MPOs), and prove the optimality of the construction. For arbitrary two-body interaction Hamiltonians, we find that an efficient description always exists, and the required tensor dimension scales linearly with the number of particles. For interesting special cases such as nearest neighbor couplings or couplings of a fixed range, a constant bond dimensions suffices. A particular efficient description exists for systems with pairwise interaction Hamiltonians of the same kind, but with arbitrary inhomogeneous coupling strengths. In addition, exponentially decaying coupling strengths (see \cite{MCPV08,CDV08}) as well as polynomially increasing coupling strengths (and combinations thereof) can be efficiently described. We discuss the possibility to {\em approximate} high-dimensional MPOs by lower dimensional ones, both analytically and numerically. For Hamiltonians corresponding to polynomially decaying interaction strengths with possible additional inhomogeneity, we show that a low-dimensional accurate approximation is possible. This allows us to study systems with long-range couplings, e.g. arising from a dipole-dipole interaction. We use algorithms based on approximate matrix product operators and compare with exact results. We show with the help of several examples that even with an approximate representation of the Hamiltonians, ground states of such systems can be accurately obtained.
As a further application we demonstrate how to find accurate approximations of the unitary time evolution operator in form of an MPO. Especially systems with long-range interactions benefit from this method.

We generalize our constructions to other geometries, and show how to obtain tensor network operators for tree tensor networks and 2D tensor networks. Tree tensor network descriptions for quantum states have been considered in \cite{DRS02,SDV06,VDNDVB07}, and we discuss how an appropriate description of operators respecting the given geometry can be achieved and utilized.

For 2D geometries, we provide an explicit construction for arbitrary pairwise couplings, where for nearest neighbor Hamiltonians and Hamiltonians of constant range a constant tensor dimension suffices (see also \cite{MCPV08}). We obtain optimized constructions for long-range interaction Hamiltonians, thereby obtaining tensor dimensions depending on the fourth root of the system size. We also discuss possible advantages of using such a tensor network representation in the numerical algorithms.

This paper is organized as follows. In section \ref{sec:AnRe} we consider 1D chains and introduce matrix product operators. We show the explicit construction of such operators and discuss a number of special cases and examples, where we provide an optimal representation. In section \ref{sec:trunc-long-range}  we consider approximate representations of operators, and illustrate the applicability for systems with long-range interactions. In section \ref{sec:time-evolution} we show how to use MPOs for time evolution. We generalize our approach to other geometric structures in sections \ref{sec:tensor-networks} and \ref{sec:2d-networks}, and summarize and conclude in section \ref{sec:conclusion-outlook}.

\section{Notation and definitions}\label{sec:not-def}

\subsection{Matrix product states and matrix product operators}
We consider a system of $N$ particles at fixed spatial positions. Every particle has an internal degree of freedom, a ``spin'', and is described as a $d$-level quantum system. The corresponding Hilbert space is given by $\mathcal{H}=(\mathbb{C}^d)^{\otimes N}$, with dimension $d^N$ growing exponentially with the system size $N$. Quantum states are represented by state vectors $\ket{\psi}$, which can be written in the computational basis as
\begin{equation}
  \label{eq:1}
  \ket{\psi} = \sum_{i_1, \dots, i_N=1}^{d}c_{i_1,\dots,i_N} \ket{i_1}\otimes \dots \otimes \ket{i_N}.
\end{equation}
The complex numbers $c_{i_1,\dots,i_N}$ can be seen as an entries of a rank $N$ tensor $c$. In general, the description of quantum states in this form is inefficient as $d^N$ complex numbers have to be specified. Imposing a certain structure on the tensor $c$, an efficient description of the corresponding states is possible, even for large $N$. An example for such an efficient representation are the so called matrix product states (MPS), where the high-rank tensor $c$ is decomposed into a product of lower-rank tensors,
\begin{equation}
  \label{eq:2}
  c_{i_1,\dots,i_N}=\sum_{\alpha_1, \dots, \alpha_{N-1}=1}^{\chi}A^{[1]}_{i_1 \alpha_1}A^{[2]}_{\alpha_1 i_2 \alpha_2}\dots A^{[N]}_{\alpha_{N-1} i_N}.
\end{equation}
$A^{[k]}$ is related to the particle $k$, and we use square brackets to indicate that the tensor depends on the position of the particle. 
The tensors $A^{[k]}$ are of third order, except for the borders, where we have second order tensors. The index $i_{k}$ refers to the ``physical'' index, while $\alpha_{k-1}$ and $\alpha_k$ are called ``virtual'' indices. Two adjacent tensors are connected via a virtual bond of dimension $\chi$, which we will refer to as bond dimension in the following. The virtual (joint) indices are contracted (i.e. summed over) in order to obtain the tensor entries $c_{i_1,\dots,i_N}$. Notice that for fixed physical indices $i_k$, one deals with rank two tensors, i.e. matrices, and the contraction leads to a matrix product.

\begin{figure}[htbp]
\centerline{\includegraphics[width=0.75\columnwidth]{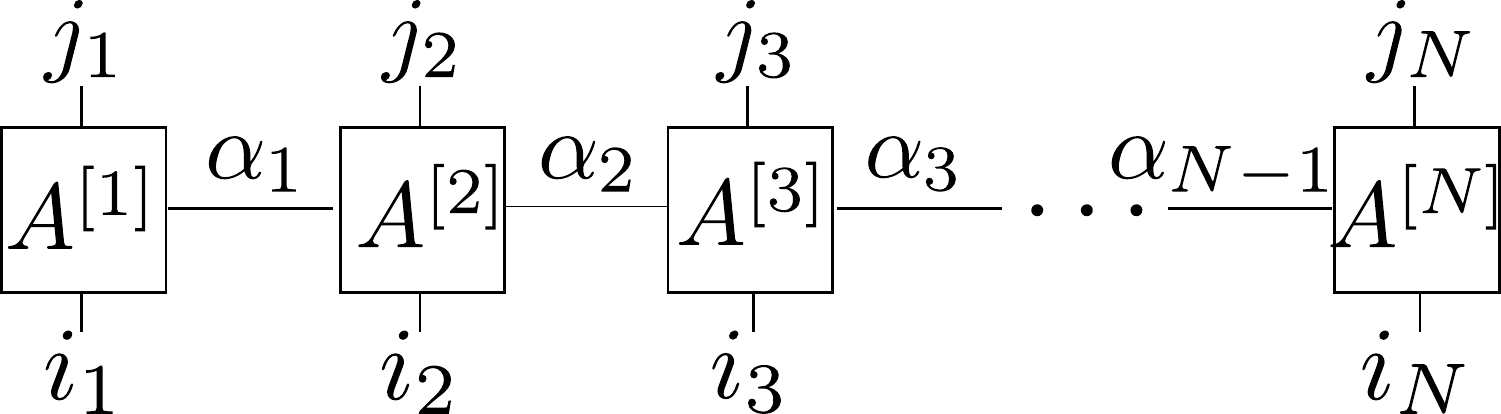}}
\caption[]{\label{fig:MPO} Matrix product operator representation. An operator $O$ acting on $N$ particles is decomposed into $N$ low-rank tensors $A^{[k]}$. Each tensor has two physical indices (input $i_k$, output $j_k$) and one or two virtual indices $\alpha_{k-1}, \alpha_k$ which are summed over.}
\end{figure}

A similar decomposition into products of low-rank tensors can also be done for operators. We consider a linear operator $O:\mathcal{H}\rightarrow \mathcal{H}$ which we decompose into basis operators $\sigma_i^j=\ketbra{i}{j}$ where $i,j=1,\dots,d$:
\begin{equation}
  \label{eq:9}
O = \sum_{\substack{i_1, \dots, i_N=1\\j_1, \dots, j_N=1}}^{d}c^{j_{1},\dots,j_N}_{i_1,\dots,i_N} \sigma^{j_{1}}_{i_{1}} \otimes \dots \otimes \sigma^{j_{N}}_{i_{N}}.
\end{equation}
We obtain a  matrix product operator (MPO) representation \cite{CB08, McC07, McC08, MCPV08, CDV08} by writing the coefficients as
\begin{equation}
  \label{eq:10}
  c^{j_{1},\dots,j_N}_{i_1,\dots,i_N}=\sum_{\alpha_1, \dots, \alpha_{N-1}=1}^{D}A^{[1]j_{1}}_{i_1 \alpha_1}A^{[2]j_{2}}_{\alpha_1 i_2 \alpha_2}\dots A^{[N]j_{N}}_{\alpha_{N-1} i_N},
\end{equation}
see figure \ref{fig:MPO}. We end up with tensors of fourth order (third order for the boundaries). Again, every tensor is related to a particle and has now two physical and two virtual indices. We write $D$ for the bond dimension of operators.

Every matrix can be written as an MPO but in the generic case this leads to a exponentially large bond dimension of $D=d^{N}$. Nevertheless a large set of useful operators have an efficient description. For example, MPO representations of Hamiltonians to describe nearest neighbor interactions and long-range interactions with exponentially decaying coupling constants were considered. All these MPOs have a constant bond dimension with respect to the system size.

We seek for efficient state- and operator representations because the lower the bond dimensions $\chi$ and $D$ are, the faster one can perform numerical computations of scalar products and expectation values. The latter is a central task in many variational methods, as e.g. the expectation value of the energy $\langle \psi|H|\psi\rangle$ has to repeatedly computed in order to find an optimal approximation to the ground state among a given class of states.
Given two states $\ket{\psi}$ and $\ket{\phi}$ represented by MPS with both bond dimension $\chi$ and an MPO for the operator $O$ with bond dimension $D$, the calculation of the complex number $\bra{\psi}O\ket{\phi}$ is performed by contracting the corresponding tensor network, i.e. by summing over the physical indices. The ``$j$- indices'' of the MPO in Eq. (\ref{eq:10}) are contracted with the physical indices of the state $\ket{\phi}$, the ``$i$-indices'' with the physical ones of $\ket{\psi}$. The calculation of the quantity scales as $\mathcal{O}(\chi^3Dd+\chi^2D^2d^2)$.

We finally remark that the bond dimension $\chi$ of an MPS depends on the entanglement of the state with respect to a given bi-partitions of the chain \cite{V03}. The maximal Schmidt rank of all possible Schmidt decompositions along the chain equals the lowest possible $\chi$. Similarly the bond dimension $D$ of an MPO corresponds to the maximal amount of entanglement the operator can create.

\subsection{Illustrations of matrix product operators}
In the remainder of the article we will provide explicit constructions of tensor networks for Hamiltonian operators. To this aim, it is useful to provide illustrations of fourth-order tensors, which we will do in the following.

\subsubsection{Matrix picture}
One possibility is to see four-rank tensors as matrices which entries are again matrices. The virtual indices correspond to the ``outer'' matrix, the physical ones to ``inner'' matrix, see also \cite{McC07}, i.e. $A_{\alpha_{k-1}i_k\alpha_{k}}^{[k]j_k}=(A_{i_k}^{[k]j_k})_{\alpha_{k-1}\alpha_{k}}$. As an example we consider the nearest neighbor two-body Hamiltonian
\begin{equation}
  \label{eq:7}
  H=\sum_{i=1}^{N-1}X_i\otimes Y_{i+1},
\end{equation}
where $X$ and $Y$ denote arbitrary single-particle operators. $H$ can be described by the site-independent tensors
\begin{equation}
  \label{eq:8}
  A^{[i]}\equiv A=
  \begin{pmatrix}
    \mathbbm{1}&X&0\\0&0&Y\\0&0&\mathbbm{1}
  \end{pmatrix};
\end{equation}
the boundaries have the form $A^{[1]} = (\mathbbm{1}, X, 0)$ and  $A^{[N]}=(0,Y,\mathbbm{1})^{T}$.

\subsubsection{Automata picture}\label{Sec:Automata}
We also refer to another picture for the tensors of the MPO, namely as automata which set operators on their related sites depending on the input from their left and right virtual indices, see reference \cite{CB08}.

We consider the tensor $A^{[k]}$ at site $k$ and refer to the virtual indices $\alpha_{k-1}$ and $\alpha_{k}$ as left and right input respectively. For fixed values of $\alpha_{k-1}$ and $\alpha_{k}$, the resulting object $(A^{[k]j_{k}}_{\alpha_{k-1} i_k \alpha_k})_{i_k}^{j_k}$ is an operator acting on the site $k$, where the values of the virtual indices $\alpha_{k-1}$ and $\alpha_{k}$ fix which operator appears.  Notice that in principle all combinations of virtual left- and right indices at different sites can occur, however some of them are not accepted, i.e. lead to a zero operator. Any allowed combination of left and right indices with a corresponding non-zero operator will be called a ``rule''. If we consider two connected tensors, the right input of the left tensor has to equal the left input of the right tensor, as these two tensors share this virtual index. The resulting Hamiltonian is a sum of all possible combinations of chains of inputs with the corresponding operators set at each of the sites.

One may also view a chain of tensors as follows: For a certain input, the first tensor sets an operator at site one and produces an output (right virtual index), which is at the same time the input for the next tensor. The second tensor then sets an operator at site two, and produces an output of the next virtual index and so forth. Notice that at each stage, several combinations might be possible, as for a given left input one can have different compatible rules, i.e. different values of right inputs with different corresponding operators to be set. The final Hamiltonian is then a sum of all possible combinations. For open boundary conditions, one has to fix the left input of the first tensor and the right input of the last one. Throughout this paper, we will always choose our rules in such a way that the virtual index can only increase from left to right, i.e. only rules $(i_i,i_2)$ with $i_1 \leq i_2$ occur. Hence we start with boundary condition one on the left side and end up with $D$ on the right side.

We discuss the Hamiltonian of Eq. (\ref{eq:7}) to clarify this construction. We consider the rules of table \ref{tab:rulepic}.
\begin{table}[htbp]
\vspace{4mm}
\begin{tabular}{c @{\quad}|  c c @{\quad} c}
rule-number&(left, right) input && output\\
  \hline
\textit{1}&$(1,1)$&$\rightarrow$&$\mathbbm{1}$\\
\textit{2}&$(1,2)$&$\rightarrow$&$X$\\
\textit{3}&$(2,3)$&$\rightarrow$&$Y$\\
\textit{4}&$(3,3)$&$\rightarrow$&$\mathbbm{1}$\\
\end{tabular}
\caption[]{\label{tab:rulepic} Set of rules which correspond to the Hamiltonian of Eq.(\ref{eq:7}). For every other combination of left and right input, the output operator is the zero operator.}
\end{table}
For open boundary conditions we fix the inputs at the left and right end of the chain. Here we choose $1$ at the left, and $3$ at the right end, i.e. the first tensor can only set the rules \textit{1} or \textit{2} and similar for the last one.

Note that this set of rules can be translated directly into an explicit construction to build up the tensors of the MPO. The element $(n,m,k,l)$ in a tensor is just the number $P_{kl}$ of the operator $P$ which is connected with the rule that has as left input $n$ and as the right one $m$. (Compare table \ref{tab:rulepic} with Eq. (\ref{eq:8}).) The bond dimension $D$ is given by maximal number of inputs, where in this case we have $D=3$.

\section{MPO representation for 1D quantum systems with long-range interactions}\label{sec:AnRe}

In this section we explicitly construct MPO representations for long-range interactions in 1D quantum systems. In the first part we consider generic two-body interactions and provide an explicit construction of the corresponding MPOs (see also \cite{CB08, McC07, McC08, MCPV08, CDV08}). We then discuss Hamiltonians with special symmetries and show that in these cases one can find MPOs with lower bond dimension. In Appendix \ref{sec:proof}, we show that these constructions are optimal in the sense that the resulting MPO have minimal bond dimension. Finally we consider general $k$--body interactions and discuss the construction of the corresponding MPOs.

\subsection{General two-body interactions}\label{sec:general-two-body}

In this section we consider general two-body interaction Hamiltonians. Starting from the example of the nearest neighbor interaction of equation (\ref{eq:7}) we first construct the MPO for long-range interactions of a fixed range $r$. Next we extend this construction to arbitrary interaction ranges $q \leq r$ and finally we indicate how to extend this representation to general two-body interactions. The main result of this section is that all two-body Hamiltonians can be expressed by an MPO with a bond dimension that grows at most linearly with the chain length, $D=\mathcal{O}(N)$.

In the first step we consider a Hamiltonian which consists of simple two-body interactions of fixed range $r$, i.e. only particles at a distance $r$ interact pairwise:
\begin{equation}
  \label{eq:27}
H=\sum_{i=1}^{N-r}X_i\otimes Y_{i+r}.
\end{equation}
It is straightforward to generalize the rules of table \ref{tab:rulepic} to long-range interactions of this form. Instead of $Y$ we set $\mathbbm{1}$ in rule \textit{3}, i.e. $(2,3) \rightarrow \mathbbm{1}$, and demand additional rules $(k,k+1)\rightarrow \mathbbm{1}$, for $k=3,\dots,r$. Finally we impose the rules $(r+1,r+2)\rightarrow Y$ and $(r+2,r+2)\rightarrow \mathbbm{1}$. We have now $r+2$ instead of three possible inputs, leading to a bond dimension of the resulting MPO with $D=r+2$.

Next we include all two-body interactions with a range $q \leq r$, i.e we consider a Hamiltonian of the form
\begin{equation}
H=\sum_{q=1}^{r}\sum_{i=1}^{N-q} X_i\otimes Y_{i+q}.\label{eq:3}
\end{equation}
Our starting point is the rule set for the fixed distance. We show in the following that setting additional rules for lower ranges we do not increase the bond dimension, which stays equal to $D=r+2$. We begin from the left side of the chain, where we still have the boundary condition 1. A string of identities is set by the rule number \textit{1} until a site $i$, where the output $X_i$ occurs. The input of the right side for this tensor equals therefore 2. Up to now there exists only the possibility to set $r-1$ identities while altering the right rule level until the range $r$ is reached and the operator $Y_{i+r}$ appears.

\begin{figure}[htbp]
  \centerline{\includegraphics[width=0.55\columnwidth]{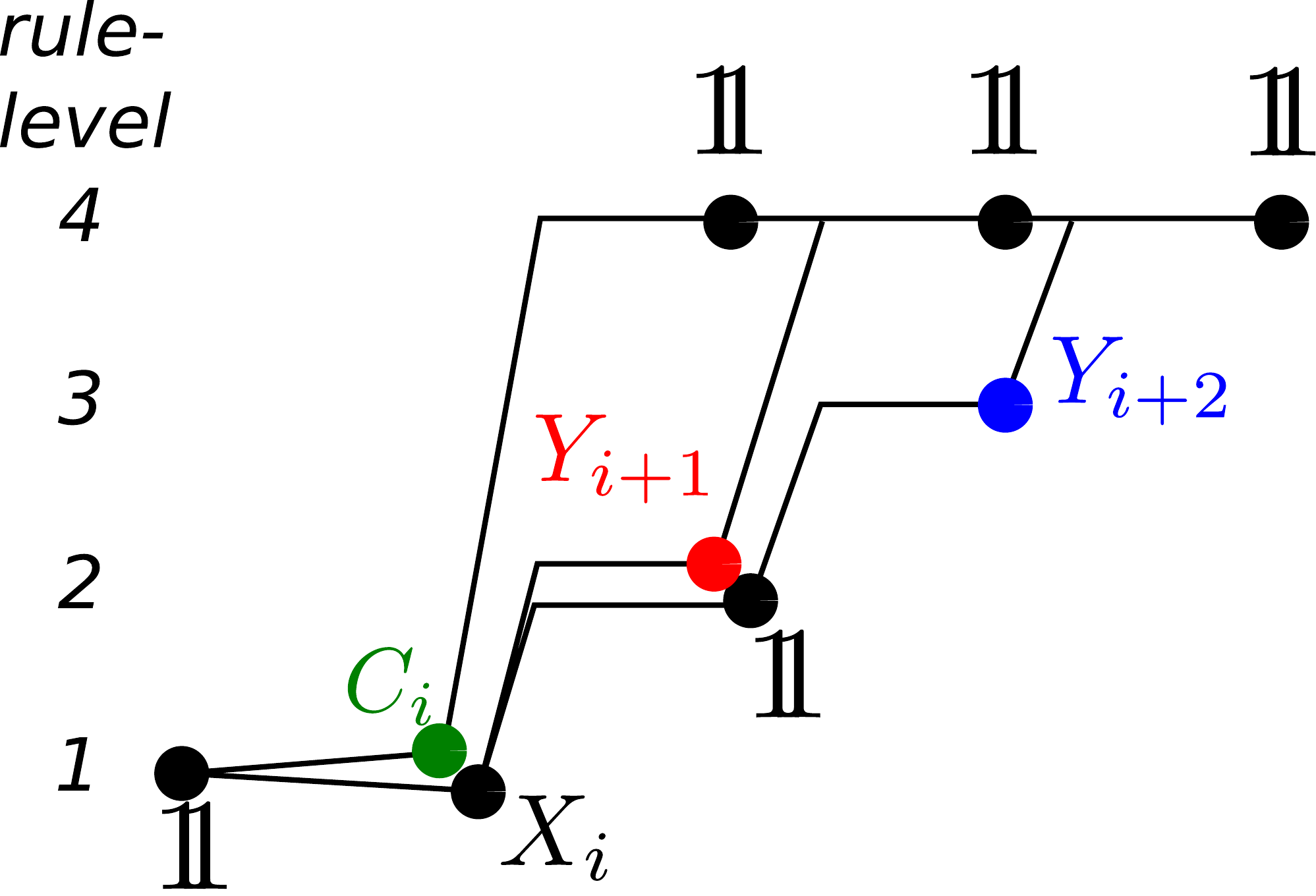}}
\caption[]{\label{fig:Embed} (Color online) Sketch of embedding local term and nearest neighbor interaction into a next-nearest neighbor Hamiltonian. There exist three possible resulting operators compatible with the set of rules: $C_{i}$, $ X_iY_{i+1}$ and $X_iY_{i+2}$.}
\end{figure}

We can demand additional rules which set the operator $Y_{i+q}$ after $q<r$ steps and lead directly to the top level, i.e. $(q+1,r+2)\rightarrow Y_{i+q}$. Doing this for all ranges smaller than $r$ we end up with a MPO which embeds all ranges without increasing the bond dimension. In addition, one can obtain a local term $C_{i}$ by adding the rule $(1,r+2)\rightarrow C$. The construction is illustrated in figure \ref{fig:Embed}.

Finally we generalize this construction to arbitrary two-body interactions, thereby going beyond the single term $X\otimes Y$ for each pairwise interactions we have discussed so far. We consider the Hamiltonian
\begin{equation}
  \label{eq:20}
  H=\sum_{i<j}h_{ij}^{[ij]},
\end{equation}
where $h_{ij}^{[ij]}$ acts non-trivially only on the sites $i$ and $j$ and can be site dependent. $h_{ij}^{[ij]}$ can always be decomposed in some basis
\begin{equation}
  \label{eq:4}
h_{ij}^{[ij]}=\sum_{k,l=1}^{d^2}\tau_{kl}^{[ij]}\sigma^k_i\otimes\sigma_{j}^{l}=\sum_{k=1}^{d^2}\sigma^k_i\otimes\tilde{\sigma}^{k [ij]}_{j}
\end{equation}
with $\tilde{\sigma}^{k [ij]}_{j}=\sum_{l=1}^{d^2}\tau^{[ij]}_{kl}\sigma_{j}^{l}$.

In our construction all ranges $q$ are realized such that we use $X_i$ for all pairs $X_iY_{i+q}$, $q=1,\dots,r$. It is thus important to shift all non-trivial information about  $h_{ij}^{[ij]}$ to the left side. In this manner we can extend the set of rules for every term in Eq. (\ref{eq:4}) such that each term can be chosen independently, i.e. with arbitrary operators and arbitrary coefficients. In the generic case the required bond dimension for a Hamiltonian of range $r$ increases to $D=d^2  r+2$.

For open boundary conditions, we have a maximal range of $N-1$ which leads to an MPO with bond dimension
\begin{equation}
D=d^2(N-1)+2,
\end{equation}
where every spin interacts with all other spins completely individually. We have therefore shown that any Hamiltonian which consists only of two-body interactions and local terms can be represented in terms of an MPO with a bond dimension that depends at most linearly on the system size. This bond dimension is optimal, i.e. there does not exist any construction which leads to a smaller bond dimension, which is proved in Appendix \ref{sec:proof}.

\subsection{Hamiltonians with symmetries}\label{sec:special-instances}
We now discuss some special cases where the Hamiltonian obeys certain symmetries or restrictions. We use the general construction described above to obtain the corresponding MPOs and show that a (significant) reduction of the required bond dimension $D$ is possible under certain circumstances. First we consider the situation where the two-body interactions are of the same kind for all pairs of particles and differ only in their strength. In this case we we can reduce the bond dimension by a factor of $1/2$. We then discuss classes of long-range interactions that can be represented by a MPO with constant bond dimension. Again, the achieved bond dimensions are optimal, see Appendix \ref{sec:proof}.

\subsubsection{Fixed type of interaction for all pairs}
\label{sec:same-inter-kindy}
In many physical systems one encounters Hamiltonians that consist of sums of identical interactions on few particles, varying only in the coupling strength, i.e. in equation (\ref{eq:20}) we have
\begin{equation}
h_{ij}^{[ij]}=c_{ij}h_{ij},\label{eq:23a}
\end{equation}
with some fixed, site independent $h_{ij}$ and arbitrary coupling strengths $c_{ij}\in\mathbbm{R}$. In this case we are able to reduce the bond dimension of the corresponding MPO by a factor of one half.

We consider a bi-partition of our system into a left part $A$ and a right part $B$ and regard the virtual bond between them as an information canal \footnote{In the momentum space DMRG \cite{X96}, one encounters non-locality of the Hamilton operator even for nearest-neighbor interaction. The interaction terms can be grouped depending on which blocks they act on. This contains already the spirit of an efficient operator representation as discussed here.}. We ask about the required information one party has to provide the other party to build up the whole Hamiltonian. Taking $h_{ij}=X_{i} \otimes Y_{j}$ as our interaction, the Hamiltonian has the form
\[H=H_A\otimes \mathbbm{1}_B+ \mathbbm{1}_A \otimes H_B + \sum_{i \in A, j \in B} c_{ij}X_i \otimes Y_j.\]
The constant $c_{ij}$ is equal to the strength of the coupling of the $i^{th}$ and $j^{th}$ particle, where $i$ lies within $A$ and $j$ within $B$. To have a complete operator, $A$ has to allocate the Hamiltonian that acts non-trivially only on $A$, the identity on $A$ and the left parts of all interactions on both $A$ and $B$. So the number of ``information-slots'' from the right site equals $2+ |A|$ (where $|A|$ denotes the number of sites in $A$). On the other hand, $B$ needs $2+ |B|$ slots.

The coupling constants $c_{ij}$ are placed into an auxiliary matrix between the two parts, which also helps to regulate the different dimensions coming from $A$ and $B$. In practice this matrix can be incorporated to the adjacent tensor with lower dimension.

A general interaction can be Schmidt-decomposed with a Schmidt-coefficient $\chi\leq d^{2}$, where $d$ equals the physical dimension per site. The bond dimension between any two tensors is hence equal to $2+\chi {\rm min}(|A|,|B|)$ and is site-dependent. The maximum required bond dimension is in the middle, where ${\rm min}(|A|,|B|) = \lfloor N/2 \rfloor$. So a more efficient description of Hamiltonians with a fixed type of interaction --as compared to general Hamiltonians-- can be achieved.

To construct the MPO explicitly one can use the ``rule''-techniques from above. In Appendix \ref{sec:suppl-mater} we demonstrate the construction method for an explicit Hamiltonian which can be specialized e.g. to dipole-dipole interactions with polynomial decay of the coupling constant, which are discussed in section \ref{sec:trunc-long-range}.

\subsubsection{Interactions that can be described by MPOs with constant bond dimension}
\label{sec:polyn-with-const}
We will now discuss two-body long-range interactions with coupling constants that depend only on the relative distance between the two interacting particles. We consider a Hamiltonian of the form
\begin{equation}
  \label{eq:6}
  H=\sum_{q=1}^{N-1}\sum_{i=1}^{N-q}c_qh_{i,i+q},
\end{equation}
where $h_{i,i+q}$ has the same form for all pairs $(i,i+q)$. Notice that $q$ denotes the distance between two sites, and $c_q$ is the corresponding coupling constant.
\newline
{\em Exponentially decaying interactions:---}
As shown in \cite{MCPV08,CDV08,McC08}, one can create MPOs which represent exponential decreasing (or increasing) coupling constants with a bond dimension that is constant, i.e. does not depend on the system size. Given a real number $\beta$, the coupling strength of Eq. (\ref{eq:6}) equals $c_{q}=\beta^{q}$. In the next paragraph we extend this to periodic boundary conditions \footnote{We still use an MPO with the structure for open boundary conditions introduced in this paper. A change to the structure of periodic boundary conditions does not improve the result.}, i.e.
\begin{equation}
c_{q}=\beta^{q}+\beta^{N-q}.
\end{equation}

We first review the construction of the exponential function in table \ref{tab:expdecay}. This can be done by adding an extra rule to the rule-set for the nearest neighbor Hamiltonian of table \ref{tab:rulepic}.
\begin{table}[htbp]
\vspace{4mm}
\begin{tabular}{c @{\quad}| c c   @{\quad} c}
rule-number&(left,right) input &  & output\\
  \hline
\textit{1}&$(1,1)$&$\rightarrow$&$\mathbbm{1}$\\
\textit{2}&$(1,2)$&$\rightarrow$&$X$\\
\textit{3}&$(2,2)$&$\rightarrow$&$\beta \mathbbm{1}$\\
\textit{4}&$(2,3)$&$\rightarrow$&$\beta Y$\\
\textit{5}&$(3,3)$&$\rightarrow$&$\mathbbm{1}$\\
\end{tabular}
\caption[]{\label{tab:expdecay} Set of rules that lead to an exponential decay of the coupling constant, $c_{q}=\beta^{q}$.}
\end{table}
The third rule produces a loop and therefore an arbitrary distance between the operators $X_i$ and $Y_{i+q}$. The identities in between carry a real factor $\beta$ which leads to the exponential decaying coupling constants (if $0<\beta<1$), because $\beta$ is raised to the power of the distance.

To achieve an additional factor $\beta^{N-q}$ as required for periodic boundary conditions, we rewrite the Hamiltonian of Eq. (\ref{eq:6})
\begin{equation}
  \label{eq:11}
  H = \sum_{q=1}^{N-1}\sum_{i=1}^{N-q}(1/\beta)^q (\beta^{N/2}X_i)(\beta^{N/2}Y_{i+q}).
\end{equation}
We just have to modify the output of the rule-numbers \textit{2} to \textit{4} and combine them with the original rules of table \ref{tab:expdecay} which leads to a bond dimension of $D=4$.
The generalization to arbitrary interactions results in a bond dimension $D=2d^2+2$.
\newline
{\em Extended Taylor expansion:---}
We now consider Hamiltonians of the form Eq. (\ref{eq:6}) with distant-dependent coupling strength $c_q$ that can be written as a polynomial times an exponential function in the distance $q$,
\begin{equation}
c_{q}=\sum_{k = 0}^M b_k\, q^k \alpha_{k}^{q},\label{eq:29},
\end{equation}
where $b_k, \alpha_k \in \mathbbm{R}$. We find that such Hamiltonians have a MPO representation with bond dimension $D$ depending only on the order $M$, independent of the system size $N$, $D = \mathcal{O}(M)$. Notice that for $\alpha_k=1$ this includes the Taylor series.

In table \ref{tab:Taylor} we sketch the basic idea of the construction.
\begin{table}[ht]
\vspace{4mm}
\begin{tabular}{c @{\quad}| c  c @{\quad} c}
rule number&(left,  right) input && output\\
  \hline
\textit{1}&$(1,1)$&$\rightarrow$&$\mathbbm{1}$\\
\textit{2}&$(1,2)$&$\rightarrow$&$X$\\
\textit{3}&$(2,2)$&$\rightarrow$&$\beta \mathbbm{1}$\\
\textit{4}&$(2,3)$&$\rightarrow$&$\beta \mathbbm{1}$\\
\textit{5}&$(3,3)$&$\rightarrow$&$\beta \mathbbm{1}$\\
\textit{6}&$(3,4)$&$\rightarrow$&$\beta Y$\\
\textit{7}&$(4,4)$&$\rightarrow$&$\mathbbm{1}$\\
\end{tabular}
\caption{\label{tab:Taylor} Next-nearest neighbor interaction with additional loop rules between the non-trivial operators.}
\end{table}
The rules \textit{1}, \textit{2}, \textit{4}, \textit{6} and \textit{7} give rise to terms like $\beta^2X_iY_{i+2}$. With the additional loop-rules \textit{3} and \textit{5} we generate arbitrary distances $q$. But now there are several combinations of rules that can be fulfilled simultaneously and which yield to the same result. E.g. for $q=5$, we have the following allowed rule-sequences: (\textit{2-3-3-3-4-6}), (\textit{2-3-3-4-5-6}), (\textit{2-3-4-5-5-6}) and (\textit{2-4-5-5-5-6}). All of them have the same effect and the number of possible combinations grows linearly with $q$. So the overall coupling constant equals $c_q=q \beta^{q}$.

If we start with rule-sets for larger ranges than next-nearest neighbor (see also Eq. (\ref{eq:27})), and add loop-rules similar to \textit{3} and \textit{5}, we generate polynomial many possibilities for a fixed $X_iY_{i+q}$. The resulting coupling constant reads in general  $c_q=q^r \beta^{q}$, with $r$ from Eq. (\ref{eq:27}). One can thus perform an extended Taylor expansion (\ref{eq:29}) of an arbitrary distance function keeping constant bond dimension. Instances where those occur are powers of long-range interactions with exponential or polynomial decaying constants, see section \ref{sec:powers-two-body}.

\subsection{Many-body interactions}\label{sec:many-body-inter}
We now turn to Hamiltonians with many-body interaction terms and investigate the resulting bond dimension of the representing MPOs. A general $N$--body Hamiltonian consists of exponentially many interaction terms, and using the results of Appendix \ref{sec:proof} it is straightforward to see that an MPO describing such a generic $N$--body interaction requires an exponentially large bond dimension. Note, however, that not the number of interacting particles causes an exponential large bond dimension, but the Schmidt decomposition of each of the $k$--body interaction terms. That is, there exist many-body interactions that can be efficiently represented by an MPO. One such example is given by the  Hamiltonian
\[
H=\sum_i\sigma_1^{x}\otimes\dots\otimes\sigma_{i-1}^{x}\otimes\mathbbm{1}_{i}\otimes\sigma_{i+1}^{x}\otimes\dots\otimes\sigma_{N}^{x},
\]
which has a very simple representation. The MPO of this operator has the same structure as for a local Hamiltonian, one simply has to exchange the rules of $\sigma^x$ and $\mathbbm{1}$.

An exponential growth of the bond dimension $D$ for a generic $k$--body interaction appears also for long-range interaction. If we use once more the arguments of Appendix \ref{sec:proof} we see that the leading order in $D$ is proportional to $N^{k-1}$, which is consistent with the two-body interaction. Again, special symmetries lead to a significant reduction of the complexity and therefore of $D$.

\subsubsection{Local $k$--body interactions}
We notice that the methods discussed in the previous sections allow also for a systematic construction of MPOs for general $k$--body interaction Hamiltonians.
To be more precise, let us discuss the Hamiltonian of a generic local $k$--body interaction. By local we mean that only neighboring particles interact with each other. If we illustrate the corresponding tensors of the MPO as matrices with matrices as their entries (see Eq. (\ref{eq:8})), we get a block structure
\begin{equation}
  \label{eq:30}
  A^{[i]}=\begin{pmatrix}
    \mathbbm{1}&P^{[i]}&0&0&\cdots\\0&0&Q^{[i]}&0&\cdots\\ & & &\ddots&\\ 0 &0&0&\cdots&R^{[i]}\\0&0&0&\cdots&\mathbbm{1}
  \end{pmatrix}
\end{equation}
with $k$ blocks $P^{[i]},Q^{[i]}, \dots, R^{[i]}$. The blocks are rank four tensors. The overall bond dimension depends on the Schmidt decomposition of a single interaction term and grows in general exponentially with $k$, $D=\mathcal{O}(d^{k})$. However, for certain many-body interactions a low-dimensional Schmidt decomposition exists, e.g. if each of the terms is just a tensor product of $k$ operators. The number of blocks in this decomposition depends linearly on $k$. In this case the dimension of the MPO is given by $D=k+1$.

In a similar way, one can consider non-local interactions, i.e. $k$--body interactions that take place between non-neighboring subsets of particles. This leads in general to the exponential growth previously discussed.

As an explicit example we analyze the MPO for a connected four-body interaction with terms $\sigma_z\otimes\sigma_z\otimes\sigma_z\otimes\sigma_z$,
\[
H=\sum_{k=1}^{N-4} c_{k} \sigma^{z}_{k}\otimes\sigma^{z}_{k+1}\otimes\sigma^{z}_{k+2}\otimes\sigma^{z}_{k+3}.
\]
For the representation we obtain four blocks with outer dimension one. We obtain the tensors
\begin{equation}
  \label{eq:12}
  A=
  \begin{pmatrix}
    \mathbbm{1}&c_k\sigma_{z}&0&0&0\\
    0&0&\sigma_{z}&0&0\\
    0&0&0&\sigma_{z}&0\\
    0&0&0&0&\sigma_{z}\\
    0&0&0&0&\mathbbm{1}
  \end{pmatrix}.
\end{equation}
It is straightforward to introduce site-dependent four-body interaction terms without further increasing the required bond dimension of the MPO, which is $D=5$ here.

If we insert identities times real factors on the diagonal we can also create four-body long-range interactions with exponential decreasing couplings (depending on the distances between the particles involved in the interaction), see \ref{sec:polyn-with-const}.
For other long-range behavior, more complex constructions arise.

Another example for a four-body Hamiltonian appears in the context of quantum chemistry (\cite{WM99,LS03}). This Hamiltonian describes electron-nuclei and electron-electron Coulomb interactions. To apply MPS or MPO methods, one needs to arrange the systems on a 1D chain. Therefore effective long-range interactions appear and the Hamilton representation exhibit a bond dimension that scales with $N^3$.

\section{Truncation of long-range MPOs}\label{sec:trunc-long-range}

In this section we consider the approximation of a given MPO by an MPO with lower bond dimension. We concentrate on two-body long-range interactions and investigate how well we can approximate the exact representation of an MPO of dimension $D$ --obtained by the constructions of section \ref{sec:special-instances}-- by an MPO of a given, lower bond dimension $D' < D$. We discuss two different approaches: (i) approximation of the coupling constants by sums of exponential decaying functions \cite{MCPV08,CDV08}; (ii) a numerical method.
While both methods allow a significant reduction of the bond dimension for polynomial decay of the coupling constant, we show that the numerical method is also applicable in more general situations, e.g. when dealing with inhomogeneous coupling strengths.

\subsection{Approximation of MPOs}\label{sec:approximation-mpos}
The first (analytical) method, as considered in \cite{MCPV08,CDV08}, is expressing the coupling constant of two sites by a functions which depends only on the distance $q$. We refer to this function as distance function $f(q)$. This function is approximated by sums of exponential functions, which can be represented by MPOs with constant bond dimension (see Sec. \ref{sec:polyn-with-const}).
Given $f(q)$, one has to find the coefficients $\lambda_i$ and $\beta_i$ such that the value
\begin{equation}
  \label{eq:13}
\|f(q)-\sum_{i=1}^n\lambda_i\beta_i^q\|
\end{equation}
is minimized. Here, $n$ is the number of exponential functions one uses for the approximation and in turn determines the bond dimension of the MPO. The bond dimension of the MPO is given by $\chi n+2$, where $\chi$ is the Schmidt rank of a single two-body interaction.

The second approach is a numerical procedure. With a variational Ansatz we find an MPO $\mathfrak{M}$ with a smaller bond dimension $D'$ which approximates the original MPO $M$ optimally. We stress that this algorithm is not constrained to a special kind of MPO. The numerical compression of an MPO is discussed in some more detail in the following. As a first ingredient we need a measure which allows us to judge how close the original MPO $M$ and its replacement $\mathfrak{M}$ actually
are. Given such a distance-measure, one proceeds as follows:
\begin{enumerate}
        \item Pick by random an appropriate MPO $\mathfrak{M}$ of a low bond dimension.
        \item Optimize (successively and repeatedly) each tensor of the MPO $\mathfrak{M}$
        in order to decrease the distance of $M$ and $\mathfrak{M}$.
\end{enumerate}
The crucial task is to find an efficient optimization procedure. Let
us start by looking at $M$ and $\mathfrak{M}$ as two ordinary operators
and forget their special MPO structure for a while. As distance-measure
we choose the Hilbert-Schmidt norm of the difference of
the two operators \footnote{Notice that this is equivalent to consider the Jamiolkowski Fidelity \cite{CJ-Iso} of the operators.}
\begin{equation}
        ||M-\mathfrak{M}||\text{\texttwosuperior}=\mbox{\ensuremath{\langle}M|M\ensuremath{\rangle}}+\langle\mathfrak{M}|\mathfrak{M}\rangle-2\mathfrak{Re}(\langle M|\mathfrak{M}\rangle).\label{eq:Distance}
    \end{equation}
The scalar product is given by
\[
    \langle M|\mathfrak{M}\rangle=\textrm{tr}(M^{\dagger}\mathfrak{M})=\sum_{i,j}M_{ji}^{*}\mathfrak{M}_{ij}
\]
Introducing the multi-index $m=(i,j)$ we formally write the operators
$(M_{ij})$ and $(\mathfrak{M}_{ij})$ as vectors $(M_{m})$ and $(\mathfrak{M}_{m})$
which turns their scalar product into standard scalar product for
vectors
\[
\langle (M_{ij})|(\mathfrak{M}_{ij})\rangle=\langle (M_{m})|(\mathfrak{M}_{m})\rangle
\]
This simple mapping from operators to vectors guides us in dealing
with the MPOs. By joining the two physical indices of each tensor
of the MPO in one multi-index we map an MPO onto an MPS. The task
of optimizing an MPS is already a standard procedure (see reference \cite{VCM08}
for a good review).

The optimization is essentially done by maximizing the overlap $\langle M|\mathfrak{M}\rangle$.
This might seem a little bit astonishing since the right side of equation
(\ref{eq:Distance}) indicates that the distance of $M$ and $\mathfrak{M}$
also depends on $\langle\mathfrak{M}|\mathfrak{M}\rangle$ (meanwhile
$\langle M|M\rangle=\textrm{const}$).
However, by making use of the QR-decomposition, one can ensure that the maximization procedure always results in
$\langle\mathfrak{M}|\mathfrak{M}\rangle=1$.
Every matrix $A$ can be written as
$A=Q\cdot R$ with $Q^{\dagger}Q=\mathbbm{1}$. We apply
this decomposition successively to the MPS $\mathfrak{M}$ regarding
its tensors as matrices with multi-indices. Starting from the borders
and multiplying the $R$-matrices into the yet not decomposed neighboring
tensors, we bring the MPS in the form
\begin{equation}
  \begin{split}
\mathfrak{M}=&\sum_{\substack{\alpha_{1}\dots\alpha_{N-1}\\i_1,\dots,i_N}}Q_{i_{1}\alpha_{1}}^{[1]}Q_{i_{2}\alpha_{1}\alpha_{2}}^{[2]}\dots  \mathfrak{A}_{i_{j}\alpha_{j-1}\alpha_{j}}^{[j]}\dots\\ &    Q_{i_{N-1}\alpha_{N-2}\alpha_{N-1}}^{[N-1]}Q_{i_{N}\alpha_{N-1}}^{[N]} \sigma_{i_1}\otimes\dots\otimes\sigma_{i_{N}}.
  \end{split}
\label{eq:Q-MPS}
\end{equation}
Since we do the QR-decomposition successively coming from the left
and right border there is one tensor $(\mathfrak{A}_{i_{j}\alpha_{j-1}\alpha_{j}}^{[j]})$
in the middle which is not subjected to the decomposition. This is
the tensor we are going to optimize. For all the Q-tensors we have
\begin{eqnarray*}
    \sum_{\alpha_{k-1}i_{k}}Q_{(\alpha_{k-1}i_{k})}^{\dagger[k]\widetilde{\alpha}_{k}}Q_{(\alpha_{k-1}i_{k})}^{[k]\alpha_{k}}=\mathbbm{1}^{\widetilde{\alpha}_{k}\alpha_{k}} & \textrm{for} & k<j\\
    \sum_{\alpha_{k}i_{k}}Q_{(\alpha_{k}i_{k})}^{\dagger[k]\widetilde{\alpha}_{k-1}}Q_{(\alpha_{k}i_{k})}^{[k]\alpha_{k-1}}=\mathbbm{1}^{\widetilde{\alpha}_{k-1}\alpha_{k-1}} & \textrm{for} & k>j
    \end{eqnarray*}
which results in
\[
    \langle\mathfrak{M}|\mathfrak{M}\rangle=\sum_{i_j \alpha_{j-1}\alpha_j} \mathfrak{A}_{i_{j}\alpha_{j-1}\alpha_{j}}^{*[j]}\mathfrak{A}_{i_{j}\alpha_{j-1}\alpha_{j}}^{[j]}.
\]
In other words: as long as we take care that our optimization produces
a normalized tensor $(\mathfrak{A}_{i_{j}\alpha_{j-1}\alpha_{j}}^{[j]})$
the whole MPS $\mathfrak{M}$ is normalized. Having done this procedure
the correct optimization of $(\mathfrak{A}_{i_{j}\alpha_{j-1}\alpha_{j}}^{[j]})$ consists in the already mentioned maximization of the overlap $\langle M|\mathfrak{M}\rangle$.
Since the tensor $(\mathfrak{A}_{i_{j}\alpha_{j-1}\alpha_{j}}^{[j]})$
enters only linearly in the scalar product, we can rewrite this expression
as
\begin{equation}
    \langle M|\mathfrak{M}\rangle=\sum_{i_j \alpha_{j-1}\alpha_j} C^{*}_{i_{j}\alpha_{j-1}\alpha_{j}}\cdot\mathfrak{A}^{[j]}_{i_{j}\alpha_{j-1}\alpha_{j}}=\langle C|\mathfrak{A^{[j]}}\rangle,\label{eq:28}
  \end{equation}
  where $C^{*}$ is the tensor obtained by contracting all tensors of the network $\langle M|\mathfrak{M}\rangle$ but $\mathfrak{A^{[j]}}$.
Setting
\[
    |\mathfrak{A}\rangle=\frac{|C\rangle}{\langle C|C\rangle}
\]
maximizes $\langle M|\mathfrak{M}\rangle$ under the condition $\langle\mathfrak{M}|\mathfrak{M}\rangle=1$
which is what we were looking for.

We demonstrate the applicability of the methods for a long-range Hamiltonian and calculate the ground state and the ground state energy. To this end we use a variational ansatz for MPOs, similarly as in \cite{CB08}. Although this computation already has an error, we refer to them as ``exact'' ground state and ground state energy, respectively. We expect the errors to be negligible, see the caption of figure \ref{fig:Dipole} for the estimated errors. Next we calculate the approximated MPOs for different values of the truncation parameter. We evaluate three quantities: The Hilbert-Schmidt distance between the original and the approximated MPO, the fidelity of the ground states and the relative difference between the ground energies in both cases of exact and approximated Hamiltonian.

The systems we have tested are the following:
(i) We consider Rydberg atoms loaded in a 1D optical lattice potential, which is described by a Hubbard model of Rydberg excitations \cite{S08,G94}. The corresponding Hamiltonian has a power law decay for the coupling constants,
\begin{equation}
  \label{eq:5}
  H=\Omega \sum_{j=1}^N(r_j+r_j^{\dag}) +\delta \sum_{j=1}^N n_j+\sum_{j<k}\frac{\beta_{0}}{(k-j)^3}n_jn_{k},
\end{equation}
where $r_j^{(\dag)}$ are the creation (annihilation) operators of excitations and $n_j$ is the number operator. The effective Rabi-frequency is denoted by $\Omega$, $\delta$ parametrize the detuning of the laser and finally $\beta_0/(k-j)^3$ is the strength of the dipole-dipole-interaction of the atoms and follows a cubic decay. This Hamiltonian includes already some assumptions on the special realization of the experiment, see \cite{S08} and references therein, especially \cite{G94} for the theoretical background.
(ii) In addition, we have investigated a slightly modified Hamiltonian of the same kind, where we considered random (but fixed) fluctuations of the relative positions of the sites. Hence we have also some randomness for the coupling constants. (iii) Finally we consider a long-range Ising model where the coupling constants are normally distributed, a so-called spin glass.

\subsection{Hubbard model with regular positions}\label{sec:rydberg-excitations}
\begin{figure}[htbp]
  \centerline{\includegraphics[width=1.\columnwidth]{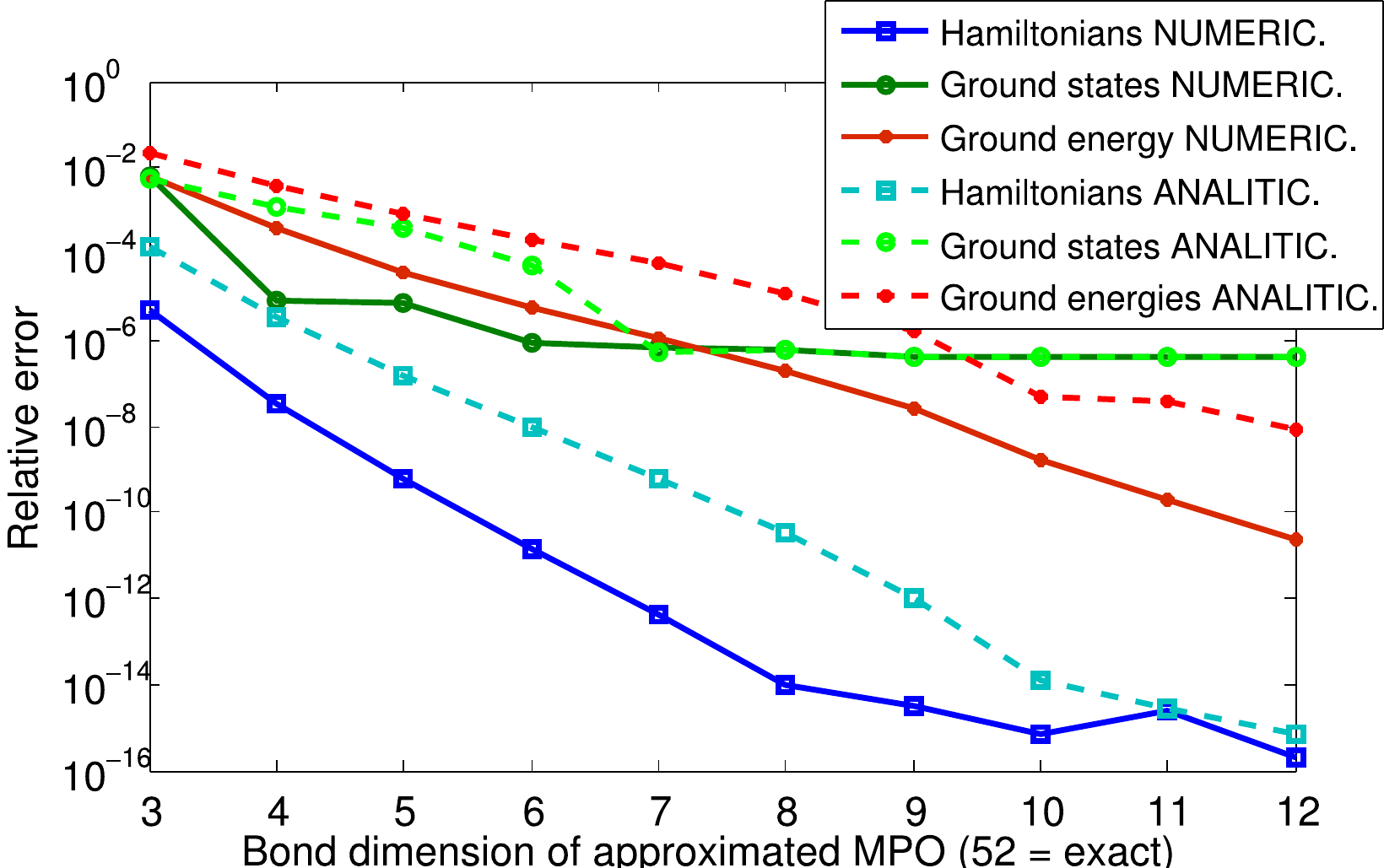}}
  \caption[]{\label{fig:Dipole}(Color online) On the quality of the MPO approximation for the Hubbard  model~(\ref{eq:5}). We choose the following parameters: $N=100$ particles (i.e. bond dimension 52 for the exact MPO); $\beta=1$, $\Omega=0.1$ and $\delta=0$. The estimated errors are: $10^{-15}$ for the operator overlap, $10^{-10}$ for the ground state energy and $10^{-10}$ for the ground state fidelity. The bond dimension of the ground state is equal to 80. Dashed lines correspond to (i) the approximation of the operator by a sum of exponentially decaying functions, while solid lines correspond to (ii) the MPO obtained by numerical truncation. Relative errors for Hamiltonian (blue), ground state fidelity (green) and ground state energy (red) as a function of the bond dimension of the approximating MPO are plotted.}
\end{figure}

We first consider a system of Rydberg atoms arranged regularly on a line, which is described by the Hamiltonian Eq. (\ref{eq:5}). Similarly as in previous works (\cite{MCPV08,CDV08}), we find that a few exponential functions suffice to describe the Hamiltonian accurately. Here we took one to ten functions which lead to a bond dimension of three to twelve, as the Schmidt-rank of a single two-body interaction is one. We have also tested the numerical optimization of the MPO approximation of the Hamiltonian. As shown in figure \ref{fig:Dipole} we find that both methods lead to accurate results, where the numerical truncation works slightly better.

Using the variational Ansatz one observes a convergence of the distance of the approximated MPO for bond dimensions larger than nine, however the error of the ground state energy still decreases for increasing bond dimension. The reason for this lies in the way of evaluation the distance between the original $M$ and approximated MPO $\mathfrak{M}$, see equation (\ref{eq:Distance}). No matter how close $M$ and $\mathfrak{M}$ are, after the division through the norms of the single operators, the scalar product is of the order $1+ \mathcal{O}(10^{-16})$, due to the rounding errors at computer precision. The outcome for the distance-measure between $M$ and $\mathfrak{M}$ is at least in the order of $\mathcal{O}(10^{-16})$. Our algorithm is capable of further reducing the the error for the ground state energy for bond dimensions between nine and twelve, although this is not visible in the operator precision. 

\subsection{Hubbard model with inhomogeneous positions}\label{sec:rand-post-hubb}
We now turn to (ii), Rydberg atoms with randomized positions. The system is still described by the Hamiltonian Eq. (\ref{eq:5}), where we consider now randomized locations $x_j=j+\sigma r_j$. Here, $r_j$ is a normally distributed random number and $0<\sigma<1$. The coupling constant of the two-body interaction equals now $\frac{\beta_0}{(x_k-x_j)^3}$. This means that the interaction strength does not show a regular decay anymore.

\begin{figure}[htbp]
  \centerline{\includegraphics[width=1.\columnwidth]{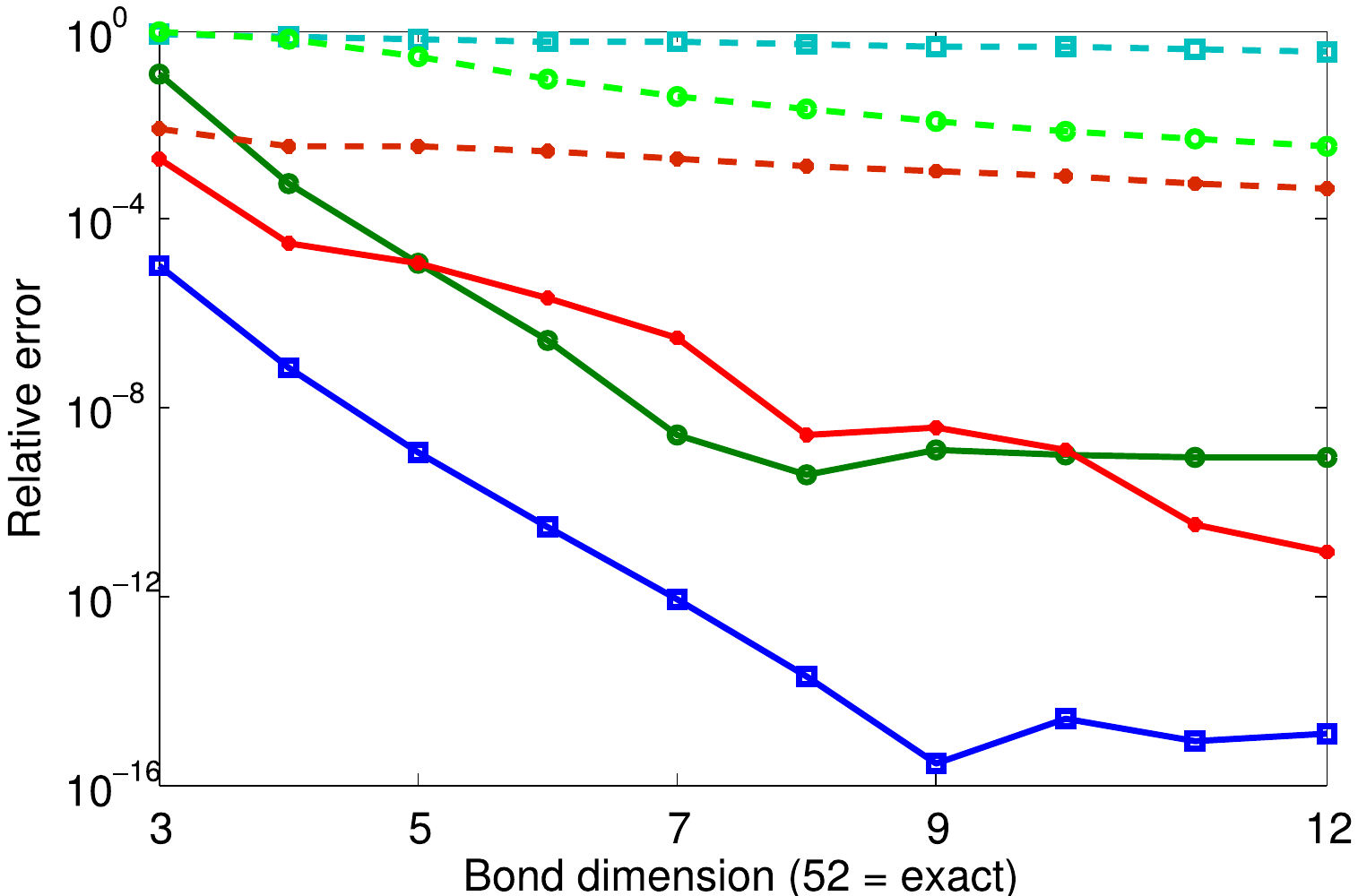}}
  \caption[]{\label{fig:DipoleMOD}(Color online) Same situation as in \ref{fig:Dipole}, except that the positions of the single particles are shifted away from the regular lattice by adding a normally-distributed number with variance 0.2 in units of the lattice distance.}
\end{figure}

It turns out that the (numerical) variational method still allows for an accurate approximation of the Hamiltonian by a MPO, where the results are as good as for the regular case (i). The method based on sums of exponential functions has to be modified to handle the new situation (see below). The achievable accuracy is significantly lower in this case, as can be seen in Fig. \ref{fig:DipoleMOD}.

We briefly discuss some adjustments of the approximation method based on sums of exponential functions. We model an irregular exponential decay with a coupling constant that depends on the absolute position of the sites: $c_{jk}=\beta^{x(k)-x(j)}$. If we change in the $i^{th}$ tensor $\beta$ to $\beta^{x_i-x_{i-1}}$, then we end up with the desired coupling constant. This can also be done for sums of exponential functions, but this special approximation of $\frac{\beta_0}{(x_k-x_j)^3}$ faces a problem: The approximated function oscillates quite heavily around the polynomial decay for $x\approx 1$ and has relatively large errors for small fluctuations at 1 but exactly at distance 1 the error almost vanishes. So in the end, the errors that occur here can be decreased, as figure \ref{fig:DipoleMOD} shows, but the method can not keep up with the numeric truncation. In particular, the precision can not be increased significantly by a higher number of exponential functions. A number of further refinements are possible, e.g. correction of nearest or next-nearest neighbor interaction terms by increasing the bond dimension of the MPO by one or two, but have not been studied in detail as the variational method already leads to an accurate result.

\subsection{Spin glass}\label{sec:spin-glass}

We finally turn to a system with completely random couplings between all pairs of particles, i.e. to a spin glass.
The Hamilton operator
\[
H=\sum_{j<k} J_{jk}\sigma_j^{z}\sigma_k^{z} + \sum_{j=1}^N B \sigma_j^x
\]
has random couplings $J_{jk}$ which follow a normal distribution. For stability reasons of the ground state algorithm we took a smaller particle number, $N=30$, and repeated the calculations several times with a negligible variance in the outcome. The bond dimension of the exact representation of the MPO is 17.

\begin{figure}[htbp]
  \centerline{\includegraphics[width=1.\columnwidth]{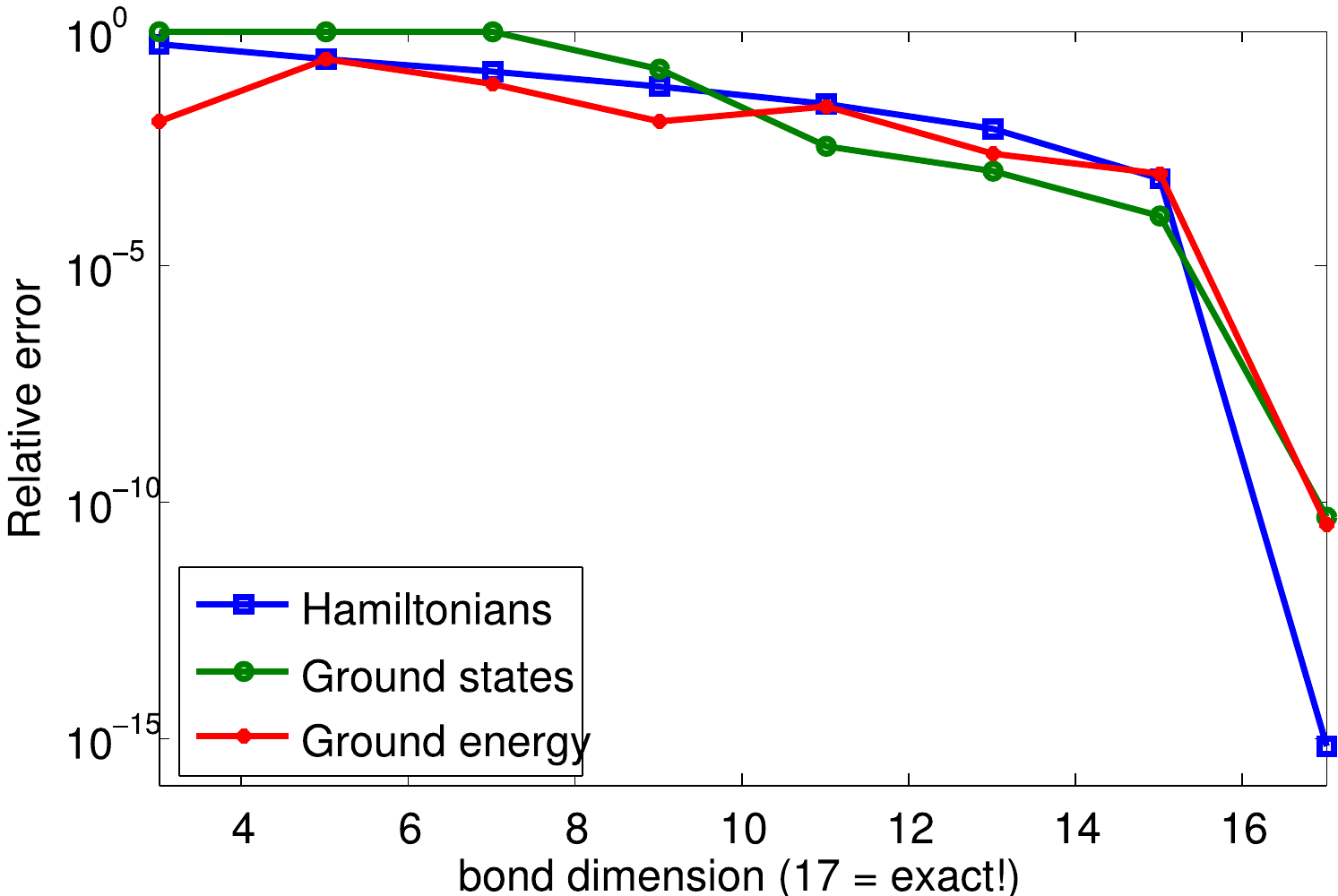}}
  \caption[]{\label{fig:spinglass}(Color online) Long-range Ising with random-couplings and transverse magnetic field with $B=1$. There is no chance to truncate an MPO such that the ground state properties are conserved. The expected errors are the same as for the first example. N=30. The bond dimension of the ground state is 80.}
\end{figure}

Using the numerical optimization method, we observe that any truncation of the operator produces an error which is at least of the order of $10^{-3}$ for the energy. Hence we conclude that an MPO of a spin glass Hamiltonian is not compressible and the full complexity is needed (see Fig. (\ref{fig:spinglass})). We find a similar result when using an approximation by sums of exponential functions

{\em Special instances of spin glasses:}---
Note that there exist special instances of spin glass realizations with a compact description in terms of an MPO. To this aim, we consider the construction of long-range exponential decaying couplings and replace the constant $\beta$ in each tensor by an independent random number. In this way we also generate instances of a spin glass, but obtain a bond dimension of the MPO which is constant. In this MPO the number of parameters is linear with the system size $N$, but $N^2$ coefficients are needed. Hence we can only generate a subset of all possible configurations. 
It is important to take into account that the distribution of the coupling constants in general cannot be carried over to the distribution of the $\beta$, as only joint probability distributions that arise from products of individual probability distributions can be described in this way. Nevertheless, for particular instances of spin glasses a compact description of the Hamiltonian in terms of an MPO is possible, leading to a significant simplification in the numerical treatment of this (subset of) cases.

\section{Time evolution with MPOs}
\label{sec:time-evolution}

As a further demonstration of the usefulness of MPOs combined with
the numerical approximation routines explained in Sec. \ref{sec:trunc-long-range} we show a way how to calculate the time evolution operator $U(\Delta t)=\exp(-iH\Delta t)$. We stress that this method includes Hamiltonians with long-range interactions. Since the time evolution operator mediates a proliferation of entanglement we are usually forced to restrict ourselves to small values of $\Delta t$. Apart from some irregularities the bond dimension needed for an appropriate MPO approximation of $U(\Delta t)$ should decrease with decreasing $\Delta t$. We are interested in a special instance of this statement: If $\Delta t$ is chosen in such a fashion that an MPO approximation of $U(\Delta t)$ with moderate bond dimension $D$ exists, the MPO approximation of any $U(\nicefrac{\Delta t}{2^{n}})$ for $n=1,2,3,...$ should also be feasible and become even easier with increasing $n$.

We focus on $U(\nicefrac{\Delta t}{2^{n}})$ because it provides the key for practical calculations. Different approximation schemes for $U(\nicefrac{\Delta t}{2^{n}})=\exp(\nicefrac{-iH\Delta t}{2^{n}})$ are available which all increase in precision with decreasing $\|\nicefrac{-iH\Delta t}{2^{n}}\|$. Thanks to the exponential dependence on $n$ already moderate values of $n$ enable us to construct very accurate MPOs for $U(\nicefrac{\Delta t}{2^{n}})$. Once the MPO for $U(\nicefrac{\Delta t}{2^{n}})$ is given, $n$ successive multiplications suffice to obtain a precise MPO approximation of the full operator $U(\Delta t)$ taking repeatedly advantage of
\begin{equation}
U(\frac{\Delta t}{2^{n-1}}) =U(\frac{\Delta t}{2^{n}})\cdot U(\frac{\Delta t}{2^{n}}).\label{eq:34}
\end{equation}

Here we have to multiply MPOs. The multiplication of two MPOs can be done tensor-wise in a straightforward way. Squaring an MPO in this fashion causes a squaring of the bond dimension. In order to avoid such an increase and to obtain an MPO approximation with the heralded bond dimension $\leq D$, we combine the multiplication with the numerical approximation method presented above (Sec. \ref{sec:approximation-mpos}).

As a final ingredient we need a method to build up MPO approximations of $U(\nicefrac{\Delta t}{2^{n}})=\exp(\nicefrac{-iH\Delta t}{2^{n}})$. Here we will consider the MPO-based Taylor expansion of $\exp(\nicefrac{-iH\Delta t}{2^{n}})\thickapprox\sum_{k=0}^{m}\nicefrac{(-iH\Delta t\cdot2^{-n})^{k}}{k!}$
with a suitable cutoff $m$. Using the Horner algorithm we get
\[\sum_{k=0}^{m}\frac{x^{k}}{k!}=1+\frac{x}{1}(1+\frac{x}{2}(\dots(1+\frac{x}{m-1}(1+\frac{x}{m}))\dots)).\]
Starting on the right side and setting $x=\nicefrac{-iH\Delta t}{2^{n}}$
we can successively build up the MPO. Calculating very precise high order approximations poses no problem when we resort to this scheme. All we need is an MPO representation of the Hamiltonian and the ability to add and multiply MPOs. Similar to the multiplication the addition of two MPOs can be done tensor-wise which results in a new MPO whose tensors have a block structure -- each block representing one of the addends. In the case of $\textrm{MPO}_{\textrm{New}}=1\!\!1+\textrm{MPO}_{\textrm{Old}}$ for each of the $N$ tensors $A^{[K]}$, $K=1\dots N$ of $\textrm{MPO}_{\textrm{New}}$
we get
\begin{eqnarray*}
A_{i,j;1,1}^{[K]\textrm{ New}} & = & 1\!\!1_{i,j}\\
A_{i,j;(\alpha+1),(\beta+1)}^{[K]\textrm{ New}} & = & A_{i,j;\alpha,\beta}^{[K]\textrm{ Old}}
\end{eqnarray*}
where $i,j$ represent the physical indices and $\alpha,\beta$ the
virtual indices.

We remark that recently a similar method has been independently introduced and utilized in \cite{SW10}.

\subsection{Test on the quality of the Taylor series}
\label{sec:test-quality-taylor}

To test the presented method, we use two different approaches. We take very small system sizes, where all objects can be calculated exactly. We have chosen $N=12$, since this allows us not only to compare the time evolved states in vector and MPS representations but additionally the unitary operator in the matrix and MPO representation. Secondly, we investigated how well the norm and the energy expectation value are conserved during time evolution of large systems. For nearest neighbor interaction, both tests can be compared with the Suzuki-Trotter decomposition of the time evolution operator, which is constructed out of products of exactly calculable exponential terms of sub-sums of the Hamiltonian. Here we used an approach taken from \cite{SS99}, which corresponds to a fourth-order Trotter decomposition. Additionally we performed also for this method the successive time doubling of a small time step (\ref{eq:34}).

The models we have considered are the XXZ-model \begin{equation}
  \label{eq:31}
H =  \cos\theta\sum_{k=1}^{N-1}\sigma^x_k\sigma^x_{k+1}+\sigma^{y}_k\sigma^y_{k+1}+ \Delta \sigma^{z}_k\sigma^z_{k+1} + \sin\theta\sum_{k=1}^N \sigma^z_k
\end{equation}
and the Bose Hubbard model Eq.(\ref{eq:5}).

\begin{figure}[htbp]
  \centerline{\includegraphics[width=1.\columnwidth]{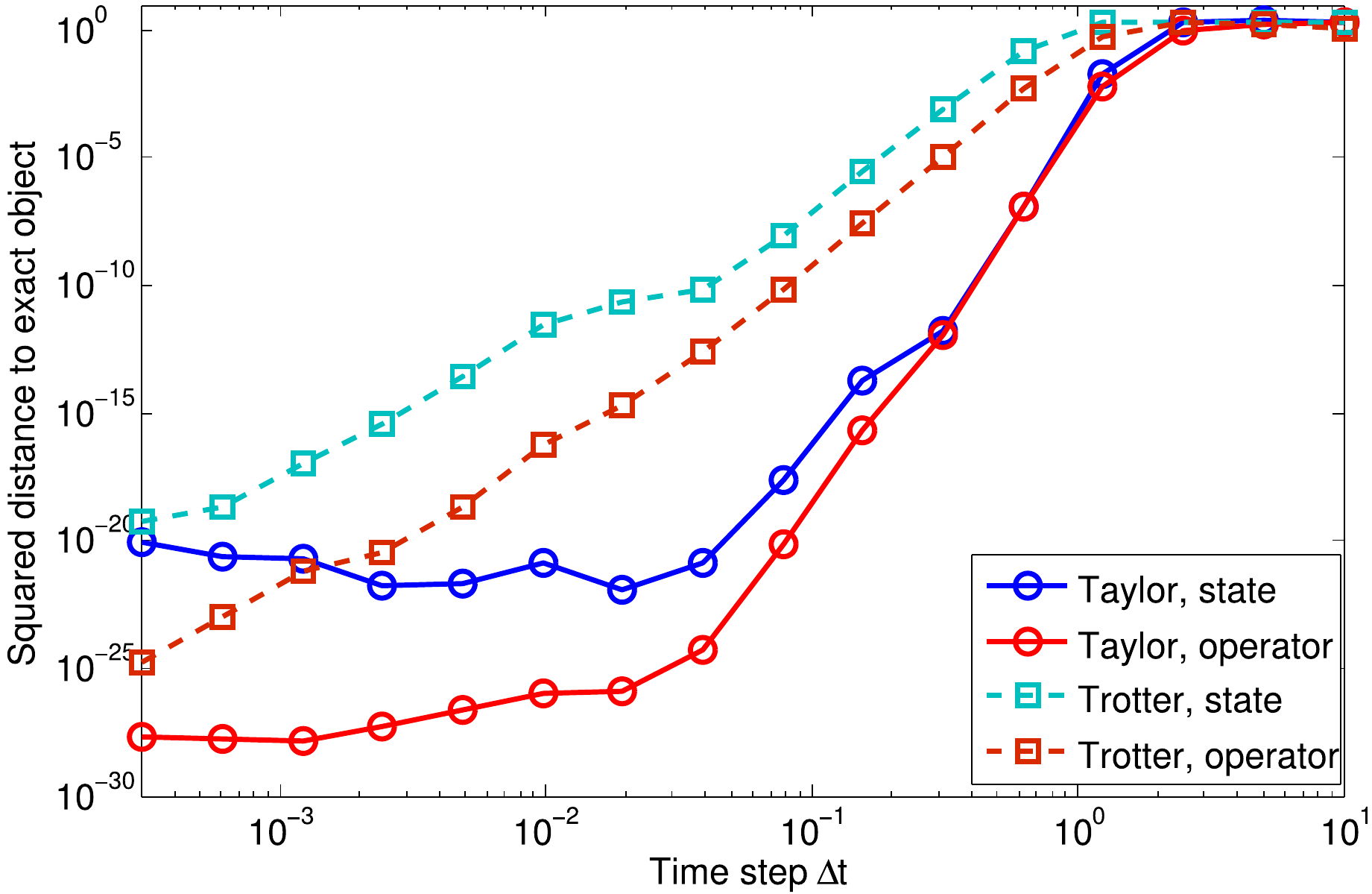}}
\caption[]{\label{fig:Small-Sys} (Color online) Comparison of Taylor expansion and Trotter decomposition for the XXZ-model on a small system, $N=12$. The squared distance between exact operator to the MPO are compared for different minimal time steps $\Delta t$ for both MPO-generating methods Taylor and Trotter. In addition, the squared distance of an exactly evolved state to the evolved MPS is measured at $t=10$ in appropriate units. The demonstrated model is defined in Eq. (\ref{eq:5}), with the parameters $\Theta = 0.35$ and $\Delta = 0.1$. The order for the Taylor series and the number of time doubling steps are adjusted for different $\Delta t$, see \cite{MvL03} for a guideline. The MPS exhibits maximal bond dimension and starts with all spins up. The bond dimension of the MPO $D_{\rm MPO}$ is restricted to 30. The Trotter method was performed with the same parameters.}
\end{figure}

The results for the first test are shown in Fig. \ref{fig:Small-Sys} for the XXZ-model. One sees that the time evolution based on the Taylor expansion method leads for small systems to better accuracies than the Trotter method, which is due to the fact that for larger $\Delta t$ a higher order of the Taylor series can be used. Furthermore we can deal easily with long-range interactions and achieve similar accuracies.

A possible drawback when using a Taylor expansion is that the approximated evolution operator is not unitary and therefore leads to errors during the time evolution. Our tests of norm and energy conservation for larger systems --here 100 particles-- show the contrary. For the XXZ-model, in fact the norm was better preserved by the Taylor series, whereas the energy deviations were exactly equally for the Taylor and Trotter method. This indicates that using MPOs --combined with successive time doubling (\ref{eq:34})-- enables us to produce faithful representations of the time evolution operator. 
\subsection{Time evolution of inhomogeneous long-range interactions}
\label{sec:time-evol-inhom}

\begin{figure}[htbp]
  \centerline{\includegraphics[width=1.\columnwidth]{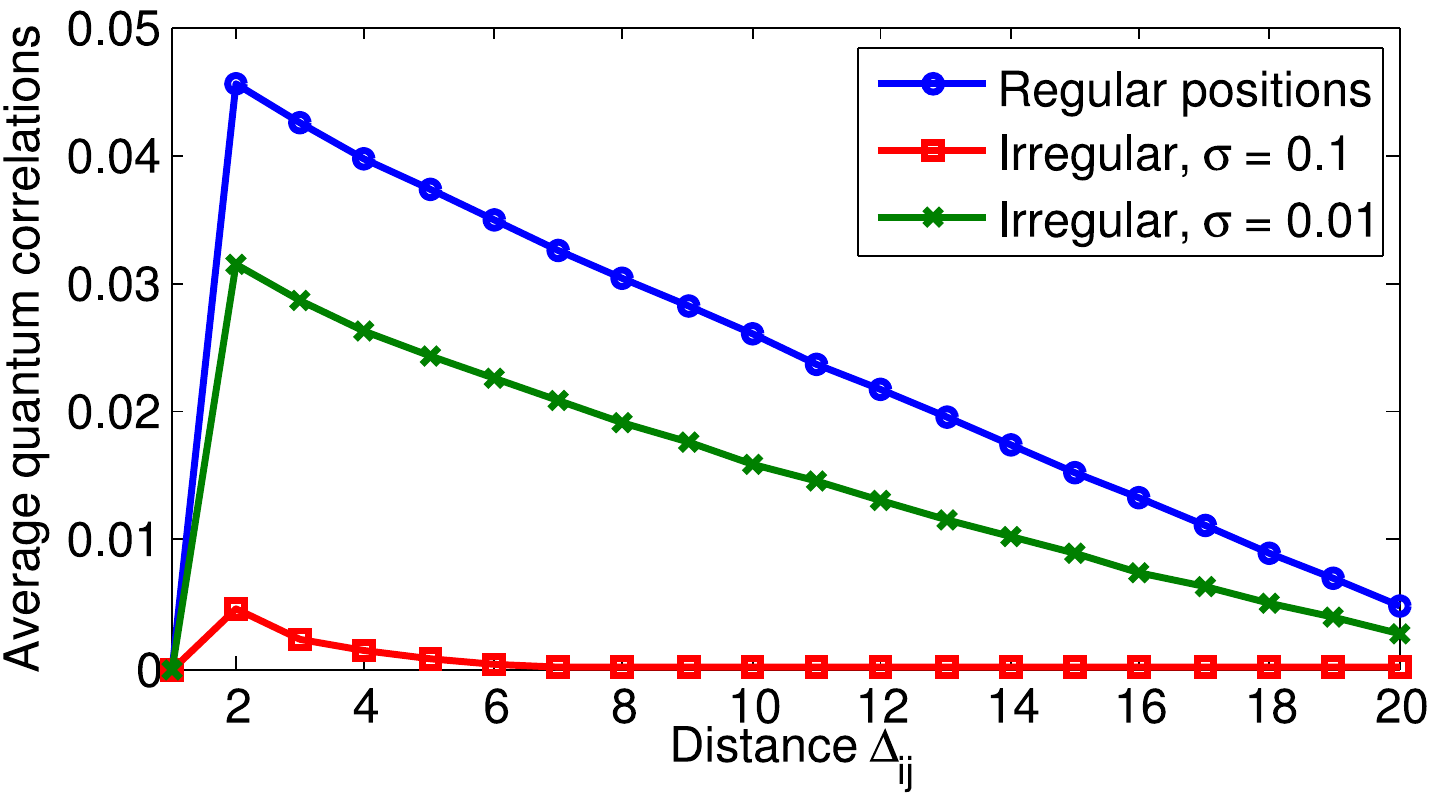}}
\caption[]{\label{fig:TimeEvol} (Color online) Two-point quantum correlation of Eq. (\ref{eq:32}). For a chain of 100 particles the averaged correlations of the sites 40 to 60 were calculated. This snapshot was taken at  $t=20$. The parameters of the model were $\beta=10, \Omega=1$ and $\delta=-12$. The unitary time evolution operator were generated by a seventh-order Taylor polynomial, subsequently five times doubled to obtain $\Delta t=0.025$; $D_{\rm MPO}=50$. We repeated the calculations for $D_{\rm MPS}=90,110,130$. The differences in the results for different $D_{\rm MPS}$ are negligible for our demonstration.}
\end{figure}

As already emphasized, expressing the time evolution operator in terms of a Taylor series takes the advantage of a simplified treatment of long-range interactions and inhomogeneous coupling strengths. We present here an example which we already discussed in the content of truncation of long-range MPOs,  namely the Hubbard-model with dipole-dipole interactions (Eq. (\ref{eq:5})). Starting with a state where all sites are in the ground state, non-classical long-range correlations between the excitations of the sites $i$ and $j$,
\begin{equation}
  \label{eq:32}
  c_{ij} = \EV{n_i\otimes n_j}{\psi(t)}-\EV{n_i}{\psi(t)}\EV{n_j}{\psi(t)}
\end{equation}
are built up for certain parameter settings. 
In the following we investigate the appearance of such long-range two-point correlations when considering inhomogeneous particle positions. As in Sec. \ref{sec:rand-post-hubb}, we randomize the positions of the atoms by adding a small, normally distributed number with a standard deviation $\sigma$. We observe in Fig. \ref{fig:TimeEvol} that a deviation of the chain positions in the order of one percent leads qualitatively to the same correlations, but if the inhomogeneity becomes larger ($\sigma=0.1$), long-range correlations are suppressed.

\subsection{Powers of two-body Hamiltonians}
\label{sec:powers-two-body}

Powers of two-body Hamiltonians $H^{n}$ are implicitly used in the construction of a small time step. Naively, one could expect an exponential growth of the bond dimension $D$ with $n$. However, in this paragraph we present several examples of short- and long-range Hamiltonians where for small $n$ an efficient representation exists. This fits well with our observations that power series can be used in the construction of the time evolution operator.

We discuss in the following the powers of the Hamiltonian describing the 1D Ising model with transverse field
\begin{equation*}
  \label{eq:15}
H = -\sum_{k=1}^{N-1}\sigma^{z}_{k}\sigma^{z}_{i+1}- B \sum_{k=1}^N \sigma_k^x.
\end{equation*}
By analyzing $H^n$ for $n=2,3,4$ analytically, we find a moderate growth of the bond dimension (see table \ref{tab:powers}). We extended this analysis to higher $n$ and determined the bond dimensions $D$ numerically by making use of an iterative procedure. Given an MPO representation for $H^{n-1}$ and $H$ we use the algorithm of Sec. \ref{sec:approximation-mpos} that finds the MPO representation of $H^{n}$ for a fixed bond dimension $D_{\rm cut}$ that is as close as possible to $H^{n-1}H$. If we start with a small $D_{\rm cut}$ and record the distance of the approximated $H^n$ to $H^{n-1}H$, we consider the exact bond dimension of $H^n$ to be found when a significant change of the distance from a finite
value to computer precision is observed.

We are able to identify the bond dimensions of the powers of the Ising model up to $n=12$. For $n \leq 4$ the numerical and analytical results are identical. For this range the bond dimension of $H^{n}$ grows much slower than the ``worst-case'' $3^{n}$. The details of our investigations are summarized in table \ref{tab:powers}.

Also the powers of the Hamiltonian of the XXZ-model of Eq. (\ref{eq:31}) have been investigated. Qualitatively the same behavior reveals, although the complexity of the model is also reflected in the powers of the Hamiltonian, see table \ref{tab:powers}.
\begin{table}[ht]
\vspace{4mm}
\begin{tabular}{l |llllllllllll}
  n & 1& 2 & 3 & 4 & 5& 6 & 7 & 8 & 9 & 10 & 11 & 12\\
  \hline
  Ising&3$^{*}$& 5$^{*}$&8$^{*}$ &12$^{*}$ & 17 & 23 &30&39&50 &64&78&97 \\
  XXZ&5$^{*}$&9$^{*}$&16 &32&51&79 &110&& &&&\\
    \end{tabular}
  \caption{Bond dimensions which have been found numerically for MPOs representing $H^{n}$ of the short-range models. $^*$Verified analytically.}
  \label{tab:powers}
\end{table}
Similarly long-range Hamiltonians with interactions that decay exponentially or polynomially with the distance exhibit efficient approximate representations of their powers. Although the exact representations of $H^n$ is high dimensional, we can find good approximations even for small bond dimensions. As an example, again we have studied the Hubbard-model for Rydberg atoms of Eq. (\ref{eq:5}), where the coupling constant decays cubically. To reach accuracies of the approximations at computer precision, we observe that for low powers $n\leq 8$ high bond dimensions are required, whereas for higher powers it is similar to the Ising model, probably since long-range terms become irrelevant due to their fast decay.

\section{General one dimensional networks}\label{sec:tensor-networks}

In this section we investigate one-dimensional generalizations of the linear tensor networks which we have discussed so far. We call a linear tensor network a set of tensors which are connected via bonds, i.e. summation over common indices, which do no form non-local loops as we encounter in two or more dimensional lattices. Examples are tree tensor networks (TTNs) \cite{SDV06} or Bethe lattices.

We distinguish between two different kinds of networks. The first type is represented by networks where every tensor carries a physical index and belongs therefore to a particle or a mode \cite{DRS02}. The second kind are networks where some of the tensors are virtual in the sense that they do not correspond to a physical particle but only have an auxiliary function \cite{SDV06}. Here we are going to introduce and discuss representations of tensor network operators (TNO) of both types and their efficient contraction. For the rest of this section we will regard only two-body interactions, however $k$--body interactions can be treated in a similar way.

\begin{figure}[htbp]
  \centerline{\includegraphics[width=0.7\columnwidth]{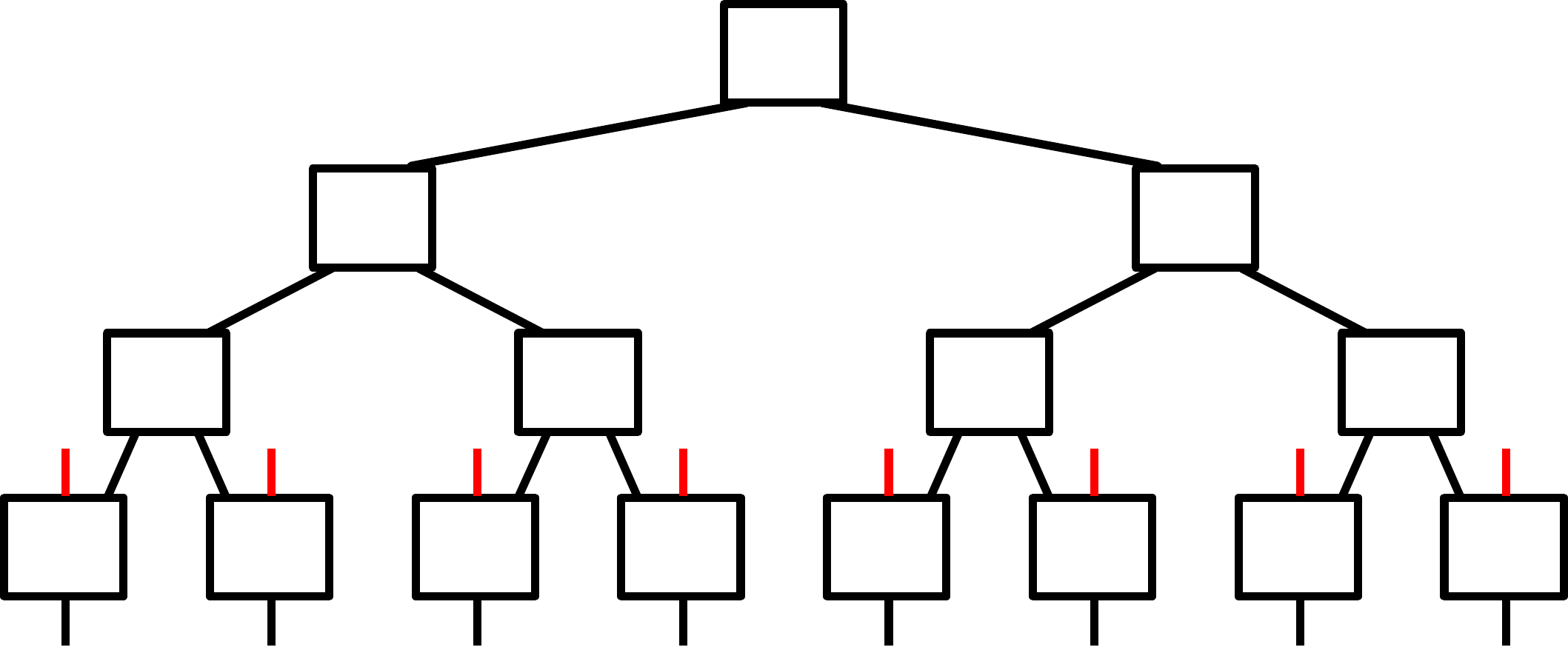}}
  \caption[]{\label{fig:TTN} (Color online) Example for the definition of a tree tensor network operator on the basis of a given definition for a tree tensor network state. Boxes indicate tensors and the red lines correspond to the new open indices, while joint indices are contracted.}
\end{figure}

Given a representation of a state in terms of a specific 1D network, we can easily define the corresponding TNO representation by increasing the order of all ``physical'' tensors by one. This additional index is open, has the same dimension as the other physical index and hence transforms the state to an operator, see figure \ref{fig:TTN}.

\subsection{Tensor networks on tree graphs}\label{sec:bethe-lattice}
We start with the first kind of networks, where all tensors represent a physical entity. It is used e.g. to mimic a geometric structure in space, see e.g. \cite{DRS02}. We consider an operator $O$ and assume that we can represent it efficiently in terms of a TNO.  As long as the contraction of two vectors in form of a scalar product $\braket{\psi}{\phi}$ is efficient, the contraction of $\bra{\psi}O\ket{\phi}$ is also efficient.

The bond dimension of TNOs representing Hamiltonians behaves very similarly to the dimension of an MPO. Nearest neighbor interaction also exhibit $D=\chi+2$.

For long-range interaction the required tensor dimension depends on the distance dependence of the coupling constants. We discuss as an instance the Cayley tree \cite{Baxter}, a finite version of the Bethe lattice where every site has the same number of neighbors without loops, except the tensors of the boundary which exhibit only one connection. We identify a center from which all the branches start. One way of modeling the coupling strength is the exponential decay of the constants. The distance between two particles is determined by the number of edges which connect them. Then a constant bond dimension can be achieved similar as explained in Sec. \ref{sec:polyn-with-const}.

We now relax this constraint on the coupling strengths and consider interactions that still depend on the distance between the two interacting sites, but be completely arbitrary in any other respect. This leads to virtual bonds of logarithmically scaling dimension. We consider the amount of information a connected sub-network $A$ has to provide to the rest of the network $B$. As in section \ref{sec:same-inter-kindy} we regard the Hamiltonian
\[H=H_A\otimes \mathbbm{1}_B+ \mathbbm{1}_A \otimes H_B + \sum_{i \in A, j \in B} c_{ij}X_i \otimes Y_j. \]
Apart from the operators $H_{A}$ and $\mathbbm{1}_{A}$ we have to allocate one ``slot'' in the information canal for every relative distance which is possible. This number grows logarithmically with the number of particles inside $A$. The maximal bond dimension equals $\chi L +2$, $L = \mathcal{O}(\log N)$ denoting the maximal distance from the cut $A-B$ within $A$.

The most general situation are arbitrary interactions for any two sites in the network. Then we have again a linear growth of the bond dimension with the system size, as we discussed in section \ref{sec:general-two-body}.

\subsection{Tree tensor networks with virtual tensors}
In the second kind of tensor networks we consider, not all tensors have a physical meaning, e.g. in the tree tensor network TTN. A TTN state is represented by Cayley tree where only the boundary tensors carry an open index. A sketch of the idea can be found in figure \ref{fig:TTN}. The additional red bars convert the object from a state to an operator.

\begin{figure}[htbp]
  \centerline{\includegraphics[width= 1.\columnwidth]{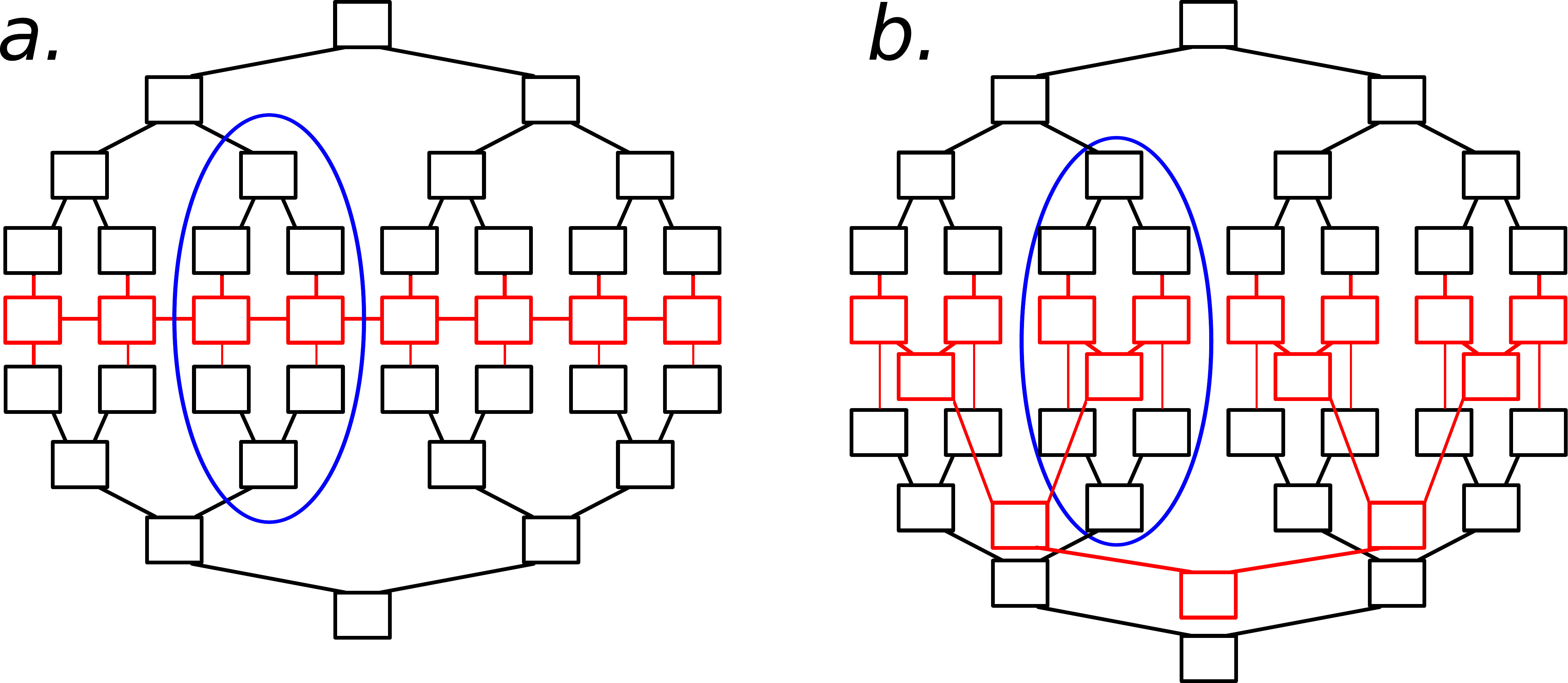}}
\caption[]{\label{fig:contraction} (Color online) The network of $\bra{\psi}O\ket{\phi}$, where $O$ is represented as (a) an MPO or as (b) a TNO. The red tensors belong to the operator, the black tensors to the vectors. The blue contour envelops the area which is contracted in the first step. The contraction using a TNO is more efficient, since the resulting tensor is of a lower rank.}
\end{figure}

We explain why the contraction of TTN states with operators in between is optimal if we have an efficient operator representation of the same structure as for the states. To this end we consider the calculation of $\bra{\psi}O\ket{\phi}$. If we insert an MPO between two TTN states and start contracting the network from the middle (see figure \ref{fig:contraction} a), we end up with tensors of higher rank. In other words, the resulting network has more loops which results into an increased effort for the contraction.

On the other hand, if we take for the operator a network-structure that is identical to the states, in the contraction tensors of lower rank appear, see figure \ref{fig:contraction} b. Because the computational speed depends polynomially on the rank of the tensors, the contraction is more efficient if the operator respects the same tensor structure as the initial state. Denoting the bond dimension of the operator representations with (a) $D_{\rm MPO}$ and (b) $D_{\rm TNO}$ respectively, the overhead caused by using the MPO is $\nicefrac{D^3_{\rm MPO}}{D^2_{\rm TNO}}$. Similarly as for states \cite{RAGE}, tree tensor networks allow for a more efficient representation of certain kinds of (long-range) interactions as compared to MPO representations.

\section{2D tensor networks}\label{sec:2d-networks}

In this section we consider two-dimensional square-networks of size
$N\times N$ and representations of operators on those systems. We take
the 2D version of the matrix product state, the so called projected entangled-pair state, PEPS \cite{VC04,MVC07}, and define the corresponding projected entangled-pair operator (PEPO), see \cite{MCPV08}. We then show how to construct explicitly PEPOs for long-range interactions. After that we discuss whether PEPOs can be used in order to improve numerical calculations, which is not equally self-evident as in the 1D case.

\begin{figure}[htbp]
\centerline{\includegraphics[width= 0.7\columnwidth]{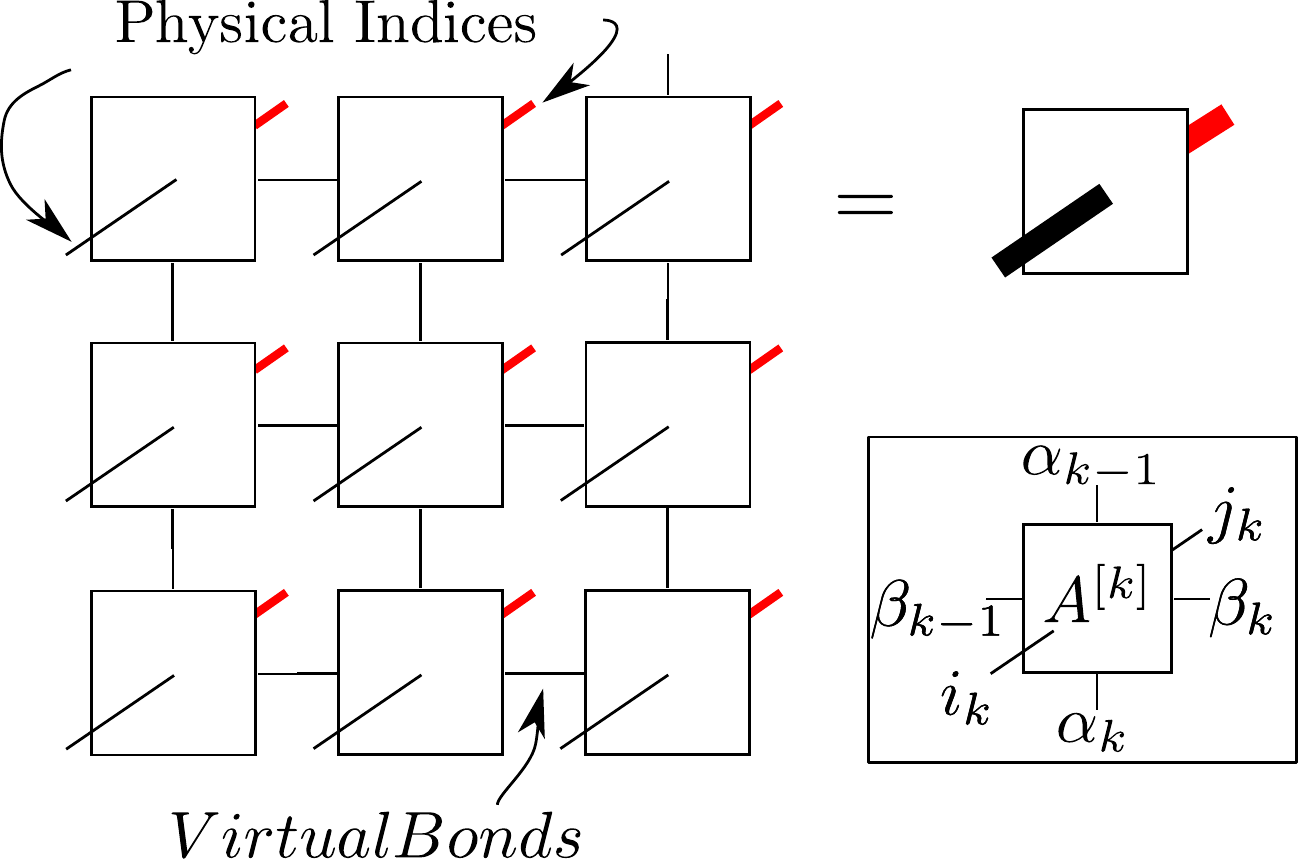}}
\caption[]{\label{fig:pepo} (Color online) On the definition of a PEPO. Without the red bars, the network represents a state; adding them, one obtains an operator. The contraction leads to the standard notation. In the insert the indices are marked as an illustration for equation (\ref{eq:14}).}
\end{figure}

A PEPS represents a state and is described by tensors of the
fifth order (expect for the borders) arranged on a 2D lattice. Four indices are connected to
neighbor tensors and hence are virtual. The fifth index is open and is
called the physical index, see figure \ref{fig:pepo}. To get the
standard notation of the vector one has to contract the network,
i.e. sum over all virtual bonds and multi-index the open indices.
Notice that this contraction is in general a numerically hard problem (NP-hard).

We define a PEPO in the
same manner as we did in section \ref{sec:tensor-networks} for 1D tensor networks. We take a
PEPS and increase the order of every tensor by one, leave this new
index open and obtain therefore two physical indices per tensor, which
correspond to an operator. Again, the contraction $\mathcal{C}$ over
all virtual bonds lead to the ``common'' matrix-notation. We decompose a
given operator in matrix form (\ref{eq:9}) into the computational basis. The coefficient for every basis-operator is then defined as
\begin{equation}
  \label{eq:14}
  c_{i_{1},\dots,i_N}^{j_1,\dots,j_N} = \mathcal{C}[\{A^{[k]i_kj_k}_{\alpha_{k-1},\alpha_k,\beta_{k-1},\beta_k}\}_{k=1,\dots,N}],
\end{equation}
which stands for a contraction over all tensors $A^{[k]}$, see also the insert of figure \ref{fig:pepo} for the definition of the indices.

\subsection{Long-range interactions}\label{sec:long-range-inter}

We have seen that in general one-dimensional operator representations for long-range interactions exhibit a bond dimension depending linearly on the system size. We show here that on a two-dimensional square-lattice the bond dimension grows like the fourth root of the system size, i.e. the square root of the side length of the grid $D\sim\sqrt{N}$.

We start by describing a less efficient representation, where the bond
dimension grows linearly with the side length of the lattice $D\sim N$. The Hamiltonian we consider is of the form
\begin{equation}
H = \sum_{i<j}c_{ij} X_i \otimes Y_j.\label{eq:25}
\end{equation}
The real coefficients $c_{ij}$ can be chosen arbitrarily. The numbering of the sum is such that we start in the upper left corner of the grid and go on to the right side. In the next row again we begin at the left side. So given a $X_i$, $Y_j$ occurs either to the right of $X_i$ or anywhere below it \footnote{To obtain a symmetric operation, replace $Y$ by $X$. Here we use $Y$ to avoid confusion.}.

\begin{figure}[htbp]
\centerline{\includegraphics[width= 0.7\columnwidth]{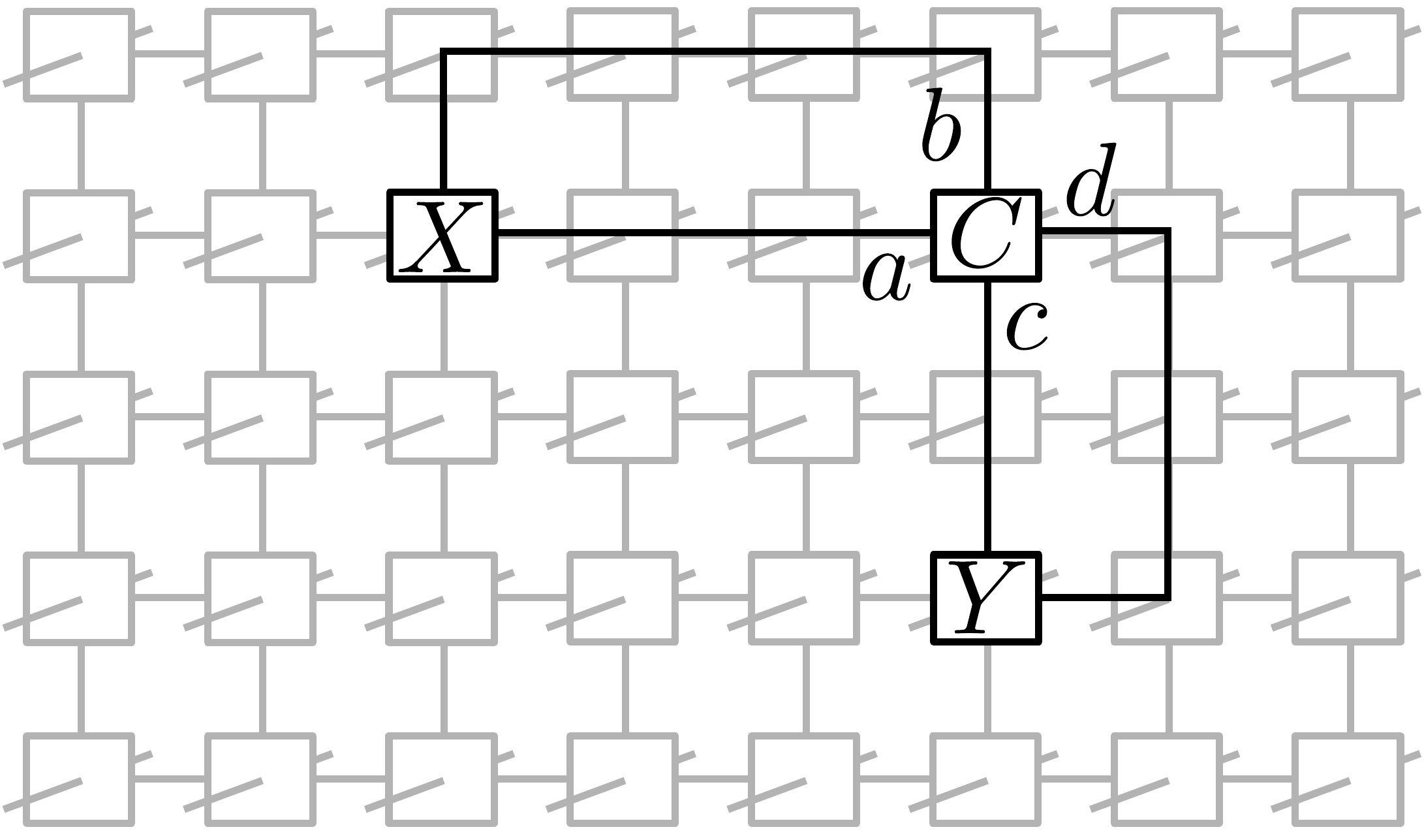}}
\caption[]{\label{fig:LRSQ} A sketch on how the information of the relative positions of $X$ and $Y$ are carried to the tensor $C$ in an optimal way.}
\end{figure}

For any individual interaction pair, the coefficient $c_{ij}$ can be provided by any tensor of the network, a good choice is the tensor which is the intersection point of the horizontal line through $X_{i}$ and the vertical line through $Y_j$, see also figure \ref{fig:LRSQ} for an illustration. We name this tensor in the following coefficient-tensor $C$. The tensors in the direct line of $C$ and $X_{i}$ and $Y_{j}$ respectively have the function to count the distance between them so $C$ ``knows'' which coefficient should appear. This is the same principle as for the 1D case. The maximal distance that can occur is $N-1$, hence the bond dimension of this construction grows linearly with the side length and we can explicitly achieve $D=N+1$.

Every tensor in a PEPO has four virtual connections to its neighbors. The actual construction uses only two of them to transport the information. This is suboptimal. In the following we will use all four inputs of $C$.
To this end we notice that every integer $m\in[1,N-1]$ can be uniquely written as $m=aL +b$ with $a,b\in[1,\sqrt{N-1}]$, with $L = \lceil \sqrt{N-1}\rceil$. So instead of transporting the information about the distance between e.g. $X_i$ and $C$ via one chain of tensors we use two parallel chains, one carrying $a$, the other $b$, see figure \ref{fig:LRSQ}.

We found a set of rules that leads to a PEPO representing the Hamiltonian (\ref{eq:25}) with bond dimension $D=2L+6$ of the horizontal bonds and $D=L+6$ for vertical bonds. The factor two in the first case is due the two possibilities, where $Y_j$ is to the left or to the right of $X_i$. The increased constant overhead comes from internal ``communication'' between the tensors counting $a$ and $b$. The explicit construction can be found in Appendix \ref{sec:suppl-mater2}, where all rules are listed. Notice that the scaling of the bond dimension in this construction  is optimal. We have $\mathcal{O}(N^{4})$ coefficients $c_{ij}$. The number of tensors equals $N^2$, where every tensor contains $D^4d^2$ parameters. We need at least  $D\sim\sqrt{N}$ to maintain the total number of parameters.

In a similar way, also short-range interactions of range $k$ and in particular nearest neighbor coupling can be treated and the corresponding PEPOs can be constructed. In Appendix \ref{sec:pepo-repr-near} this is explicitly done for nearest neighbor couplings. Also general two-body interactions, not only consisting of two operators $X_i \otimes Y_i$, can be treated similarly as in the 1D case.

\subsection{Are PEPOs useful?}
\label{sec:are-pepos-useful}

In this paragraph we discuss whether PEPOs can help to increase the efficiency of numerical algorithms. In many occasions one needs to calculate expectations values and scalar products. We give analytical and numerical indications under which conditions we can use PEPOs to improve computational performance.

Here we concentrate on two-body nearest neighbor and long-range interactions, described by the Hamiltonian (\ref{eq:25}) with $Y_j\equiv X_j$.
The state we consider is denoted by $\ket{\psi}$ and described by a PEPS of dimension $\chi$. We are interested in calculating the expectation value $E=\bra{\psi}H\ket{\psi}$. In order to calculate $E$ we have to sum over all indices, which is in general a hard problem. We consider therefore the approximate contraction scheme proposed in \cite{VC04,MVC07,VCM08}. As a first step we reduce this three-layer structure to a single-layer structure by summing over the physical indices, leaving us with a 2D tensor network. Next we start from the left side of this new network and replace the first two columns by a single column which is as close as possible to original ones. Mathematically, this corresponds to applying an MPO to an MPS, and approximating the resulting MPS by a lower dimensional one. 
Repeating this procedure we end up with a single column (MPS) which can be contracted efficiently with the final MPS. As the bond dimension of the new columns would grow exponentially with the number of contracted columns, we have to truncate the columns and allow only a maximal bond dimension $D_{\rm cut}$.

If we do not use a PEPO to calculate $E$, we have to contract all single terms of the Hamiltonian individually, i.e. we have to repeat the contraction of  $\bra{\psi}X_i\otimes X_j\ket{\psi}$ for all pairs $(i,j)$, which can be up to $\mathcal{O}(N^4)$ terms for a general two-body interaction. Note that the dimensions of this network remain constant compared to the network of the norm $\braket{\psi}{\psi}$.
For nearest neighbor interactions we can use also successively MPO-slices which contain the interactions of one row or one column. It is almost as resource-saving as the term-wise calculation but significantly faster. However, this method is not applicable in the case of long-range interactions.

On the other hand the calculation of $E$ using a PEPO requires only a single contraction of the 2D tensor network, but leads to some extra cost in the calculation.
Compared to a contraction of the scalar product $\braketd{\psi}$, there are two sources that slow down the calculation of $E$. First the dimension of the network after the summation over the physical indices grows from $\chi^2$ to $\chi^2D$. Secondly, because of the increased complexity of the network, the required bond dimension $D_{\rm cut}$ in the approximate contraction scheme needs to be increased in order to obtain a similar accuracy. Hence it is not clear whether and under which conditions the usage of a PEPO improves the calculation of $E$.

Using PEPOs obtained by our general construction (see Appendix \ref{sec:suppl-mater2} and \ref{sec:pepo-repr-near}), we find that the contraction for product states is efficient. This follows from the rule structure, and one can in fact show that a {\em linear} increase in the required tensor dimension $D_{\rm cut}$ allows for an exact treatment. In contrast, general 2D tensor networks, e.g. the representation of the time evolution operator of the Ising model without external field \cite{MCPV08}, lead to exponentially growing bond dimension for contraction, even though the PEPO has low dimension. 
For states with a PEPS representation with bond dimension $\chi \geq 2$, we find by numerical simulations that the usage of a PEPO requires an increased $D_{\rm cut}$ coming with higher computational costs. We also tried further PEPO representations in order to circumvent the increase of $D_{\rm cut}$. E.g. we used a general 1D comb-like structure in the spirit of Sec. \ref{sec:bethe-lattice}, which does not exhibit vertical virtual bonds except for the right-most column. Even though this PEPO does not contribute to the dimension of the vertical indices, we observed a similar increase of $D_{\rm cut}$. This can be seen as an indication that the complexity of the network causes a larger $D_{\rm cut}$ and less importantly the augmentation of the tensor dimensions.

In contrast, when calculating $\EV{\mathbbm{1}+tH}{\psi}$ with $t\ll 1$, we found that $D_{\rm cut}$ for $\EV{\mathbbm{1}+tH}{\psi}$ and $\braketd{\psi}$ are of the same order, and the computational cost using PEPOs is smaller. Similarly, for larger systems with long-range interactions PEPOs are favorable since term-wise calculations suffer an overhead of $\mathcal{O}(N^4)$.

\section{Conclusion}\label{sec:conclusion-outlook}

In this paper we have investigated tensor network operator representations for long-range interaction Hamiltonians. For general 1D systems with two-body interactions, we provided systematic, explicit constructions of MPOs with bond dimension growing only linearly with the system size. For systems respecting certain symmetries or restrictions, we have shown that a significant reduction of the bond dimension can be achieved. We also proved that the representations we obtain are optimal, i.e. have minimal bond dimension.

We have also investigated approximate representations of operators using low-dimensional MPOs based on analytical and numerical methods. We found that Hamiltonians corresponding to systems with (inhomogeneous) decaying long-range couplings can be represented with help of low-dimensional MPOs, while for systems with completely random couplings no truncation is possible.

Using such an MPO-based approach, we have discussed and investigated applications for ground-state approximation and time evolution. We demonstrated that the usage of approximate MPO representation allows for an accurate numerical treatment of certain models, including systems with (inhomogeneous) long-range interactions. In the context of time evolution we make use of effective time-doubling based on a Taylor-series approach.

Finally we have generalized our approach to other tensor network geometries, including tensor trees and 2D networks. For 2D systems, we have explicitly constructed an efficient representation for long-range interaction Hamiltonians in terms of a PEPO and discussed under which conditions PEPOs can help to increase numerical performance. 

The presented techniques and methods are applicable in ground state approximation and time evolution of strongly correlated quantum systems, where in particular a treatment of systems with long-range interactions is possible.

\begin{acknowledgments}
This work was supported by the FWF and the European Union (QICS,
SCALA,NAMEQUAM). We would like to thank the
Erwin-Schr\"odinger-Institut in Vienna for its hospitality during the \textit{Quantum
  Computation and Quantum Spin Systems} workshop in 2009, where many fruitful discussion have taken place.
\end{acknowledgments}


\appendix

\section{Example for section \ref{sec:same-inter-kindy}}
\label{sec:suppl-mater}

Here, an explicit construction of the MPO for a Hamiltonian $H=\sum_{k=1}^NX_k+\sum_{k<l}c_{kl}Z_k\otimes Z_l$, $c_{kl}\in\mathbbm{R}$ is provided. The rules are specified in the tables \ref{tab:ExpEx1} to \ref{tab:ExpEx4}. Notice that the Hamiltonian (\ref{eq:5}) describing a Hubbard model of Rydberg excitations is a special instance thereof.

To have a clear structure of rules, we insert auxiliary matrices $T^{[k]}$ between the tensors $A^{[k]}$ of the MPO, such that the coefficients of the operator (\ref{eq:9}) are
\begin{equation*}
  \label{eq:26}
  c^{j_{1},\dots,j_N}_{i_1,\dots,i_N}=A^{[1]}_{i_1j_1}T^{[1]}A^{[2]}_{i_2j_2}T^{[2]}\dots T^{[N-1]}A^{[N]}_{i_Nj_N}.
\end{equation*}
In practice the $T^{[k]}$ can be drawn into the physical tensors, $\tilde{A}^{[k]}_{i_kj_k}=A^{[k]}_{i_kj_k}T^{[k]}$. We assume an even number of particles $N$; for an odd number, some small corrections in the middle of the chain have to be made.
\begin{table}[ht]
\vspace{4mm}
\begin{tabular}{c @{\quad} |c c @{\quad} c }
rule-number&(left, right)- input&&output\\
  \hline
  \textit{1}&$(1,1)$&$\rightarrow$&$\mathbbm{1}$\\
  \textit{2}&$(1,2)$&$\rightarrow$&$Z$\\
  \textit{3}&$(1,D)$&$\rightarrow$&$X$\\
  \textit{4}&$(m,m+1)$&$\rightarrow$&$\mathbbm{1}$\\
  \textit{5}&$(D-1,D)$&$\rightarrow$&$Z$\\
  \textit{6}&$(D,D)$&$\rightarrow$&$\mathbbm{1}$\\
\end{tabular}
\caption[]{\label{tab:ExpEx1} Rules for $A^{[k]}$, The bond dimension equals $D={\rm min}(k+2,N-k-3)$; $m=2,\dots,D-2$.}
\end{table}
\begin{table}[ht]
\vspace{4mm}
\begin{tabular}{c @{\quad}| c  c @{\quad} c }
  rule-number&(left, right)- input&&output\\
  \hline
  \textit{1}&$(1,1)$&$\rightarrow$&$1$\\
  \textit{2}&$(m,m)$&$\rightarrow$&$1$\\
  \textit{3}&$(m,k+1)$&$\rightarrow$&$c_{k-m+2,k+1}$\\
  \textit{4}&$(k+2,k+3)$&$\rightarrow$&$1$\\
\end{tabular}
\caption[]{\label{tab:ExpEx2} Rules for $T^{[k]}$, $k<N/2$: The matrix dimension is equal to $k+2\times k+3$; $m=2,\dots,k+1$.}
\end{table}
\begin{table}[ht]
\vspace{4mm}
\begin{tabular}{c @{\quad}| c @{\quad} c @{\quad} c }
  rule-number&(left, right)- input&&output\\
  \hline
  \textit{1}&$(1,1)$&$\rightarrow$&$1$\\
  \textit{2}&$(2,m)$&$\rightarrow$&$c_{k,N-m+2}$\\
  \textit{3}&$(m+1,m)$&$\rightarrow$&$1$\\
  \textit{4}&$(N-k+3,N-k+2)$&$\rightarrow$&$1$\\
\end{tabular}
\caption[]{\label{tab:ExpEx3} Rules for $T^{[k]}$, $k>N/2$: The matrix dimension is equal to $N-k+3\times N-k+2$; $m=2,\dots,N-k+1$.}
\end{table}
\begin{table}[ht]
\vspace{4mm}
\begin{tabular}{c @{\quad}| c @{\quad} c @{\quad} c }
  rule-number&(left, right)- input&&output\\
  \hline
  \textit{1}&$(1,1)$&$\rightarrow$&$1$\\
  \textit{2}&$(m,n)$&$\rightarrow$&$c_{N/2-m+2,N-n+2}$\\
  \textit{3}&$(N/2+2,N/2+2)$&$\rightarrow$&$1$\\
\end{tabular}
\caption[]{\label{tab:ExpEx4}Rules for $T^{[N/2]}$: The matrix dimension is equal to $N/2+2\times N/2+2$; $m,n=2,\dots,N/2+1$.}
\end{table}

\section{Proof of optimality}\label{sec:proof}
Here we show that the constructions of long-range interactions of Sec. \ref{sec:AnRe} are optimal in the sense that there does not exist an alternative MPO representation with a lower bond dimension. We are going to prove this statement for the three cases we have considered so far: A completely general two-body interaction with interaction range $r\leq N/2$ \footnote{One may proof a similar statement for a range $r\leq N-1$, but in this case one finds a dependence on the structure of the individual interactions of equation (\ref{eq:4}). }, the case of site-independent interactions $h_{ij}^{[ij]}=c_{ij}h_{ij}$ from section \ref{sec:same-inter-kindy} and the further specialization of a exponential decay times a polynomial as discussed in section \ref{sec:polyn-with-const}.

The proof of optimality is based on the Choi-Jamiolkowski isomorphism \cite{CJ-Iso}, which relates operators with state vectors. The entanglement of the corresponding state vector is directly related to the entanglement of the operator, which in turn is related to the bond dimension when represented as an MPO. In particular, we will consider the entanglement of the state vector as measured by the Schmidt number, i.e. the number of non-zero Schmidt coefficients of the reduced density operator with respect to a given bi-partition of the system. For any given bi-partition, the Schmidt number provides a lower bound on the required bond dimension of the corresponding MPO. This follows from the fact that by applying a given operator, one can produce a state --the state corresponding to the operator via the Jamiolkowski isomorphism-- with a certain amount of entanglement. The amount of entanglement an MPO can produce is upper bounded by the bond dimension of the MPO. In order that an MPO provides a faithful representation of the given operator, it is thus required that its bond dimension is at least as big as the Schmidt number of the corresponding state vector.

To be more precise, we consider a bi-partition  $A-B$ of the system.
The Hamiltonian which we investigate is of the form
\begin{equation}\label{eq:17}
H = H_A + H_B + \sum_{k=\frac{N}{2}-r-1}^{\frac{N}{2}}\sum_{l=\frac{N}{2}+1}^{\frac{N}{2}+r}c_{kl} X_k\otimes Y_l^{[l]}.
\end{equation}
with $H_A \equiv H_A\otimes\mathbbm{1}^{\otimes \frac{N}{2}}$ and $H_B\equiv\mathbbm{1}^{\otimes \frac{N}{2}}\otimes H_B$. $Y_l^{[l]}$ means that the operator depends on the side it acts.
We consider the state vector $\ket{\phi}=\ket{\phi^{+}}^{\otimes N/2}\otimes\ket{\phi^{+}}^{\otimes N/2}\equiv \ket{\varphi}\otimes\ket{\varphi}$, which consists of $N$ pairs of the $\ket{\phi^{+}}$ Bell state. $\ket{\phi}$ is not entangled with respect to the bi-partition. The state corresponding to the operator $H$ is given by
\begin{equation}
\ket{\psi}=H\otimes\mathbbm{1}^{\otimes N}\ket{\phi},
 \end{equation}
where $H$ acts on the first particle of every entangled pair. The entanglement of $\ket{\psi}$ between $A$ and $B$ is measured by the Schmidt-rank, i.e we consider the rank $r$ of the reduced density operator $\rho_A= {\rm tr}_B(|\psi\rangle \langle \psi|)$. If the Schmidt-rank $r$ of $\ket{\psi}$ equals the bond dimension $D$ of the MPO representation of $H$, we have shown that the construction is optimal. If $r < D$, there could exist a more efficient representation.

For the Schmidt-rank we calculate the reduced density matrix of $A$. One finds
\begin{equation*}
  \label{eq:16}
  \begin{split}
\rho_A =& {\rm tr}_B \ketbrad{\psi}=H_{A}\ketbrad{\varphi}H_A + \EV{H_B}{\ket{\varphi}}\ketbrad{\varphi} +\\ & \sum_{k,k'=\frac{N}{2}-r-1}^{\frac{N}{2}} \alpha_{kk'}\,    X_k\ketbrad{\varphi}X_{k'},
  \end{split}
\end{equation*}
with
\begin{equation*}
  \label{eq:18}
\alpha_{kk'} = \sum_{l,l'=\frac{N}{2}+1}^{\frac{N}{2}+r} c_{kl}\,c_{k'l'}\,\EV{Y_l^{[l]}Y_{l'}^{[l']}}{\ket{\varphi}}.
\end{equation*}
A further summation over $k'$ in the last term leads to a density operator of the form
\begin{equation*}
  \label{eq:19}
  \rho_A = \sum_{k=0,\dots,r+1}\ketbra{x_k}{\tilde{x}_k}.
\end{equation*}

To show that the construction of the MPO is optimal, we have to check whether the rank of $\rho_A$ equals the bond dimension of the MPO used to represent $H$; ${\rm rank}(\rho_A) = {\rm min\{dim}({\rm span}(\ket{x_{k}})),{\rm dim}({\rm span} (\ket{\tilde{x}_{k}}))\}$.
The set $\{\ket{x_k}\}_{k}$ consists of the vectors
\begin{equation}\label{eq:21}
  \begin{split}
    &\{\ket{x_{0}}, \ket{x_{1}},\ket{x_{2}},\dots,\ket{x_{r+1}}  \}=\\
    &\{H_{A}\ket{\varphi}, \EV{H_B}{\ket{\varphi}}\ket{\varphi},
    X_{\frac{N}{2}-r-1} \ket{\varphi},\dots,X_{\frac{N}{2}}
    \ket{\varphi} \},
  \end{split}
\end{equation}
$\{\ket{\tilde{x}_k}\}_{k}$ equals
\begin{equation}
  \label{eq:22}
  \{H_{A}\ket{\varphi}, \ket{\varphi}, \sum_{k=2}^{r+1}\alpha_{2,k}\ket{x_{k}},\dots,\sum_{k=2}^{r+1}\alpha_{r+1,k}\ket{x_{k}} \}.
\end{equation}

It is clear that the set of equation (\ref{eq:21}) is linear independent as long as the set $\{\mathbbm{1}^{\frac{N}{2}}, H_A, X_{\frac{N}{2}-r-1},\dots,X_{\frac{N}{2}}  \}$ is linear independent, which is true for generic interactions. The second set, equation (\ref{eq:22}), is also linear independent for generic coefficients $c_{kl}$ and operators $Y^{[l]}$. The rank of $\rho_A$ is therefore $r+2$, which is also the bond dimension we found with our construction of a two-body Hamiltonian of this form, i.e the construction is optimal. The proof for the most general case of the Hamiltonian arbitrary interactions $h_{ij}^{[ij]}$ follows the same ideas, but is more lengthy. Again the result is that the bond dimension of our construction, $D=rd^2+2$, equals the rank of the reduced density matrix and hence optimal.

Now we treat the situation $r=N-1$ and $h_{ij}^{[ij]}=c_{ij}h_{ij}$. The simplified Hamiltonian for our considerations equals
\begin{equation}\label{eq:23}
H = H_A + H_B + \sum_{i=1}^{\frac{N}{2}}\sum_{j=\frac{N}{2}+1}^{N} c_{ij} X_i\otimes Y_j.
\end{equation}
With the same arguments from above we end up with similar sets of vectors like in the equations (\ref{eq:21}) and (\ref{eq:22}), but now with a cardinality of $\frac{N}{2}+2$. If we again allow general interactions of the form (\ref{eq:23a}), we obtain a rank $\frac{N}{2} \chi+2$.

Notice that for special choices of $c_{ij}$ we obtain a lower rank. Trivial examples are setting some coefficients to zero, another instance is an exponential decay of the coupling constant as discussed in Section \ref{sec:polyn-with-const}. If we have $c_{ij}=\beta^{j-i}$, the set (\ref{eq:22}) becomes
\begin{equation}
  \label{eq:24}
  \{H_{A}\ket{\varphi}, \ket{\varphi}, \beta^{-1}\ket{x}, \beta^{-2}\ket{x},\dots,\beta^{-\frac{N}{2}}\ket{x} \}
\end{equation}
with $\ket{x}=\sum_{k,k',l,l'}\beta^{l+l'-k'}\EV{X_lX_{l'}}{\ket{\varphi}}\ket{x_k}$. This set is highly linearly dependent, in fact it spans a three-dimensional space, exactly what we get for the bond dimension of the MPO. In the same way, other distance functions such as Eq. (\ref{eq:29}) can be inserted to prove the optimality of the representations.

\section{2D long-range interaction}
\label{sec:suppl-mater2}

We present in this section the explicit construction of the long-range Hamiltonian on a square lattice of section \ref{sec:long-range-inter}. We use the same picture as in the 1D case, namely the ``rule-picture''. Every tensor has four inputs (left, right, up, down) which go from one to $D$, the bond dimension. On grounds of these numbers an operator is set at the tensors site.

The tables \ref{tab:LRSQ1} to \ref{tab:LRSQ4} list the rules for the long-range interaction representation (\ref{eq:25}) for 2D lattices. In the following, $m,n,o$ and $p$ go from one to $L=\lceil\sqrt{N-1}\rceil$; when the corresponding rule-number is stared, the numbers are only from the set $[1,L-1]$.

While the rules so far had always integers from 1 to $D$, we use here  more symbolic inputs from the set $\{e,c,d,f,g,-L,\dots,L\}$. The cardinality of this set equals the bond dimension. The constant $c_{mnop}$ is the coupling constant and indicates the horizontal distance between $X$ and $Y$ with $(m-1) L+n$ and the vertical one with $(o-1) L+p$.

\begin{figure*}[thbp]
{\includegraphics[width=0.6\textwidth]{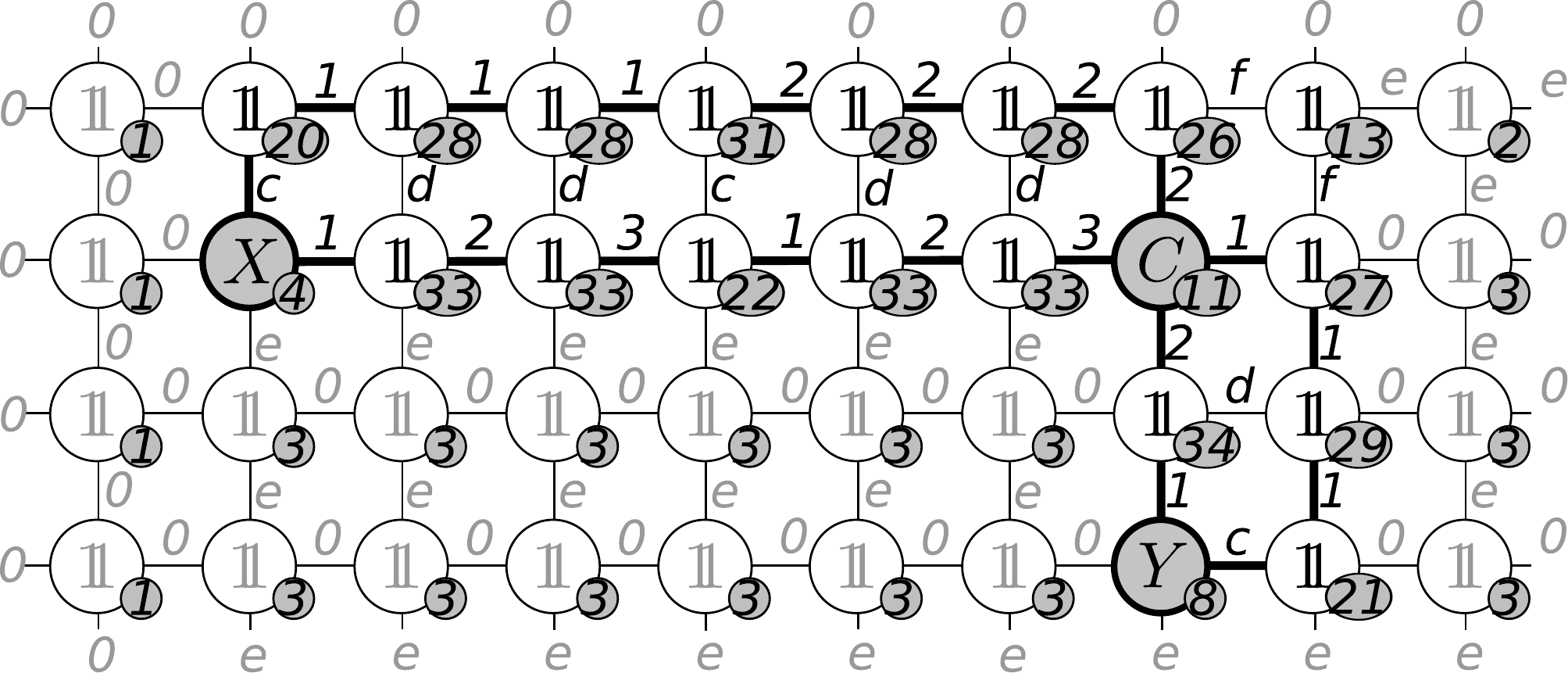}}
\caption[]{\label{fig:lrsqRules} This sketch shows the combination of rules for long-range interaction with $N=10$ for the instance that $Y$ is to the right of $X$. The gray circles show the rule number of the tables. The ``trivial'' rules are shaded and the ``communication'' lines are thicker, compare also with figure \ref{fig:LRSQ}; $C\equiv c_{2312}\mathbbm{1}$. }
\end{figure*}

An instance of the combinations is given in figure \ref{fig:lrsqRules}.

\begin{table}[th]
\vspace{4mm}
\begin{tabular}{c @{\quad}| c  c @{\quad} c }
  rule-number&(left, right, top, bottom)- input&&output\\
  \hline
  \textit{1}&$(0,0,0,0)$&$\rightarrow$&$\mathbbm{1}$\\
  \textit{2}&$(e,e,0,e)$&$\rightarrow$&$\mathbbm{1}$\\
  \textit{3}&$(0,0,e,e)$&$\rightarrow$&$\mathbbm{1}$\\
\end{tabular}
\caption[]{\label{tab:LRSQ1}``Trivial'' rules for 2D long-range construction.}
\end{table}
\begin{table}[th]
\vspace{4mm}
\begin{tabular}{c @{\quad}| c @{\quad} c @{\quad} c }
  rule-number&(left, right, top, bottom)- input&&output\\
  \hline
  \textit{4}&$(0,1,c,1)$&$\rightarrow$&$X$\\
  \textit{5}&$(m,e,g,n)$&$\rightarrow$&$C_{0-1mn}X$\\
  \textit{6}&$(-1,0,c,e)$&$\rightarrow$&$X$\\
  \textit{7}&$(0,c,s,g)$&$\rightarrow$&$X$\\
  \textit{8}&$(0,c,1,e)$&$\rightarrow$&$Y$\\
  \textit{9}&$(m,f,n,e)$&$\rightarrow$&$C_{nm00}Y$\\
  \textit{10}&$(0,g,g,e)$&$\rightarrow$&$Y$\\
\end{tabular}
\caption[]{\label{tab:LRSQ2}``Interaction'' rules for 2D long-range construction.}
\end{table}
\begin{table}[th]
\vspace{4mm}
\begin{tabular}{c @{\quad}| c c @{\quad} c }
  rule-number&(left, right, top, bottom)- input&&output\\
  \hline
  \textit{11}&$(m,n,o,p)$&$\rightarrow$&$C_{nmop}\mathbbm{1}$\\
  \textit{12}&$(m,-n,o,p)$&$\rightarrow$&$C_{-o-nmp}\mathbbm{1}$\\
  \textit{13}&$(f,e,0,f)$&$\rightarrow$&$\mathbbm{1}$\\
  \textit{14}&$(f,0,f,e)$&$\rightarrow$&$\mathbbm{1}$\\
  \textit{15}&$(0,g,0,g)$&$\rightarrow$&$\mathbbm{1}$\\
  \textit{16}&$(g,0,0,g)$&$\rightarrow$&$\mathbbm{1}$\\
  \textit{17}&$(c,e,0,c)$&$\rightarrow$&$\mathbbm{1}$\\
  \textit{18}&$(g,0,c,e)$&$\rightarrow$&$\mathbbm{1}$\\
  \textit{19}&$(m,0,c,n)$&$\rightarrow$&$C_{00mn}\mathbbm{1}$\\
\end{tabular}
\caption[]{\label{tab:LRSQ3}``Tensor $C$+surrounding'' rules for 2D long-range construction.}
\end{table}
\begin{table}[ht]
\vspace{4mm}
\begin{tabular}{c @{\quad}| c  c @{\quad} c }
  rule-number&(left, right, top, bottom)- input&& output\\
  \hline
  \textit{20}&$(0,1,0,c)$&$\rightarrow$&$\mathbbm{1}$\\
  \textit{21}&$(c,0,1,e)$&$\rightarrow$&$\mathbbm{1}$\\
  \textit{22}&$(\sqrt{N-1},1,c,e)$&$\rightarrow$&$\mathbbm{1}$\\
  \textit{23}&$(0,c,1,\sqrt{N-1})$&$\rightarrow$&$\mathbbm{1}$\\
  \textit{24}&$(-1,-\sqrt{N-1},c,e)$&$\rightarrow$&$\mathbbm{1}$\\
  \textit{25}&$(-1,e,0,c)$&$\rightarrow$&$\mathbbm{1}$\\
  \textit{26}&$(m,f,0,m)$&$\rightarrow$&$\mathbbm{1}$\\
  \textit{27}&$(m,0,f,m)$&$\rightarrow$&$\mathbbm{1}$\\
  \textit{28}&$(m,m,0,d)$&$\rightarrow$&$\mathbbm{1}$\\
  \textit{29}&$(d,0,m,m)$&$\rightarrow$&$\mathbbm{1}$\\
  \textit{20}&$(g,-m,0,m)$&$\rightarrow$&$\mathbbm{1}$\\
  \textit{31}&$(-m,-m,0,d)$&$\rightarrow$&$\mathbbm{1}$\\
  \textit{32*}&$(m,m+1,0,c)$&$\rightarrow$&$\mathbbm{1}$\\
  \textit{33*}&$(m,m+1,d,e)$&$\rightarrow$&$\mathbbm{1}$\\
  \textit{34*}&$(0,d,m+1,m)$&$\rightarrow$&$\mathbbm{1}$\\
  \textit{35*}&$(c,0,m+1,m)$&$\rightarrow$&$\mathbbm{1}$\\
  \textit{36*}&$(-m-1,-m,d,e)$&$\rightarrow$&$\mathbbm{1}$\\
  \textit{37*}&$(-m-1,-m,0,c)$&$\rightarrow$&$\mathbbm{1}$\\
\end{tabular}
\caption[]{\label{tab:LRSQ4}``Distance counting'' rules for 2D long-range construction.}
\end{table}

\section{A PEPO representing nearest neighbor interaction on a square lattice}
\label{sec:pepo-repr-near}

Here we present the PEPO we used for the numerical studies of section \ref{sec:are-pepos-useful}. Our goal is to represent the Hamiltonian
\begin{equation}
H = \sum_{<i,j>} X_i \otimes X_j.\label{eq:25xx}
\end{equation}

For the construction  we divide the virtual bonds between the tensors into two groups: ``main-bonds'' and ``auxiliary-bonds''. The bond dimension for the first kind equals three, for the latter two, note also reference \cite{M09}.
All horizontal bonds are main-bonds, whereas all vertical bonds are
the auxiliary-bonds. The only exception is that in the last column all
vertical bonds also belong to the main-class \footnote{In principle
  every vertical line can be taken. This is interesting especially if
  one performs a variational method and wants to have the
  higher-dimensional stem in the column of the varied tensor.}. The
vertical main-line we call stem, the horizontal lines are the
branches.

\begin{table}[htbp]
\vspace{4mm}
\begin{tabular}{c @{\quad}|c c @{\quad} c }
  rule-number&(left, right, top, bottom) input&&output\\
  \hline
  \textit{1}&$(1,1,1,1)$&$\rightarrow$&$\mathbbm{1}$\\
  \textit{2}&$(1,2,1,1)$&$\rightarrow$&$X$\\
  \textit{3}&$(2,3,1,1)$&$\rightarrow$&$X$\\
  \textit{4}&$(1,3,1,2)$&$\rightarrow$&$X$\\
  \textit{5}&$(1,1,2,1)$&$\rightarrow$&$X$\\
  \textit{6}&$(3,3,1,1)$&$\rightarrow$&$\mathbbm{1}$\\
\end{tabular}
\caption[]{\label{tab:2DNN} Set of rules which is needed for the branches of the PEPO representation corresponding to the Hamiltonian of Eq. (\ref{eq:25xx}).}
\end{table}

\begin{figure}[htbp]
\centerline{\includegraphics[width= 0.5\columnwidth]{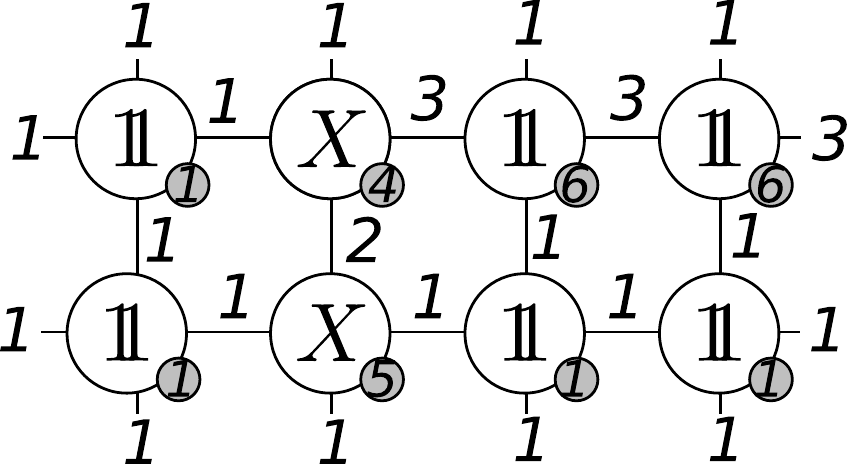}}
\caption[]{\label{fig:2DNNfinite} Detail of a possible configuration of a vertical interacting pair. The gray circles indicate the rule-number used.}
\end{figure}

We have a closer look to a single branch. The rules we use in the branch
are listed in table \ref{tab:2DNN}. We start from the left side with the left
input equal to one and move to the right side setting identities by
rule number \textit{1}. If at a certain site of a branch an $X$ occurs,
there are two possibilities. The first one is that the interaction
partner is on the right side (rule number \textit{2}) or it is the one
below it (rule number \textit{4}). In the first situation the right
input of the left partner equals two, so its right neighbor can set rule
number \textit{3} and hence has as the right input three, which is kept
up to the end of the branch (rule number \textit{6}). The situation here is identical to the MPO-case, except that now we have additional top- and bottom inputs, which are in this case fixed to one.

In the second situation the interacting particles are located one upon
the other. The upper tensor uses a rule with a bottom input of value
two, so that the lower tensor gets the signal to set rule
\textit{5}. The right input of the upper tensor already equals three, again this number is transported till the stem.

The stem has the task to coordinate all branches such that only one
interaction per addend occurs, i.e. it allows only one branch with a
right input-number three. To illustrate this construction we provide
an explicit example in figure \ref{fig:2DNNfinite}.


\end{document}